\documentclass[iop]{emulateapj}
\usepackage[backref,breaklinks,colorlinks,citecolor=blue]{hyperref}
\usepackage{lipsum}
\usepackage{float}
\usepackage{color}
\usepackage[section]{placeins}
\usepackage{epstopdf}

\shortauthors{GOSSAGE ET AL.}

\begin{document}

\title{Age Determinations of the Hyades, Praesepe, and Pleiades via MESA 
       Models with Rotation}

\author{Seth Gossage\altaffilmark{1}, Charlie Conroy\altaffilmark{1},
        Aaron Dotter\altaffilmark{1}, Jieun Choi\altaffilmark{1},
        Philip Rosenfield\altaffilmark{1}, Philip Cargile\altaffilmark{1},
        Andrew Dolphin\altaffilmark{2}
       }

\altaffiltext{1}{Harvard-Smithsonian Center for Astrophysics,
                 60 Garden Street, Cambridge, MA 02138, USA}

\altaffiltext{2}{Raytheon,
                 1151 E. Hermans Road, Tuscon, AZ 85706, USA}

\slugcomment{Accepted to ApJ}

\begin{abstract}
  The Hyades, Praesepe, and Pleiades are well studied stellar clusters that 
  anchor important secondary stellar age indicators. Recent studies have 
  shown that main sequence turn off-based ages for these clusters may 
  depend on the degree of rotation in the underlying stellar models. 
  Rotation induces structural instabilities that can enhance the chemical 
  mixing of a star, extending its fuel supply. In addition, rotation 
  introduces a modulation of the star's observed magnitude and color due to 
  the effects of gravity darkening. We aim to investigate the extent to 
  which stellar rotation affects the age determination of star clusters. We 
  utilize the MESA stellar evolution code to create models that cover a 
  range of rotation rates corresponding to $\Omega/\Omega_{\rm{c}}=0.0$ to $0.6$ 
  in $0.1$ dex steps, allowing the assessment of variations in this 
  dimension. The statistical analysis package, MATCH, is employed to derive 
  ages and metallicities by fitting our MESA models to Tycho $B_T$, $V_T$ and 
  2MASS $J$, $K_s$ color-magnitude diagrams. We find that the derived ages 
  are relatively insensitive to the effects of rotation. For the Hyades, 
  Praesepe, and Pleiades, we derive ages based on synthetic populations that 
  model a distribution of rotation rates or a fixed rate. Across each case, 
  derived ages tend to agree roughly within errors, near $680$, $590$, and 
  $110-160$ Myr for the Hyades, Praesepe, and Pleiades, respectively. These 
  ages are in agreement with Li depletion boundary-based ages and previous 
  analyses that used non-rotating isochrones. Our methods do not provide a 
  strong constraint on the metallicities of these clusters.
\end{abstract}


\section{Introduction}
\label{s:intro}
Stellar populations serve as landmarks in understanding the cosmological
timeline and as laboratories for testing stellar evolution theory. Star 
clusters are among the most powerful objects for use in calibrating stellar 
models owing to the common age, metallicity and distance of their member 
stars. The most nearby clusters are in many ways best suited for this type 
of work due to the high quality data ranging from deep color-magnitude 
diagrams (CMDs), high resolution spectroscopy, and asteroseismology. As such, 
so-called benchmark clusters like the Hyades, Praesepe, and Pleiades offer 
some of the best chances of ensuring the accuracy of our models.

Isochrone construction and usage of the Li depletion boundary (LDB) are two 
of the primary methods for accessing the ages of such clusters. 
\cite{BMG1996} were the first to successfully detect the LDB using faint 
stars in the Pleiades in conjunction with stellar models of \cite{NRC1993}, 
and derive an age estimate. Isochrones are stellar models covering a range 
of masses, paused at a moment in time, similar to how we often observe a 
collection of stars at a single moment. Hence, isochrones are an intuitive 
means of modeling stellar populations and have a long employment history in 
astronomy (e.g., \citealt{PCdSC1977,VRM2002,JL2005,EJJ2016,SXY2018}).
Both of these methods are model-dependent, simulating the observables that
we collect in databases. Therefore, these methods are subject to the adopted 
physical assumptions, which are not universal. Nonetheless, model-dependent 
age determination techniques are widely applicable. Meanwhile, they also 
serve to test our composite theory of stellar evolution, relying on all of 
its underlying framework to make credible stellar analogues and predictions. 

A number of empirical secondary age determination techniques count on 
knowing  the ages of these clusters, as derived from model-dependent methods 
(see \citealt{DS2009} for a review). For instance, gyrochronology uses these 
ages to calibrate mass-period relations (e.g. \citealt{SB2007, MH2008}) and 
offers the possibility of a high precision tool for determining ages of 
individual field stars. The zero-point of the relations between these 
quantities and age is set by the assumed ages of clusters like the Hyades,
Praesepe, and Pleiades. The accuracy of  gyrochronological results may be 
compromised if significant uncertainty exists in the ages on which its 
formalism is built.

Researchers have theorized how rotation may affect the behavior of stars for 
nearly a century (\citealt{HvZ1924}; also see e.g., \citealt{SS1929} and
references therein). However, large databases of stellar models that
incorporate these effects have only become available within the last decade
or so; e.g. STERN \cite{IB2011}, Geneva \cite{SE2012} (also 
\citealt{MeMa1997,AM1997,MaZh1998,AM1999,MaMe2000a,MeMa2000}), and recently 
MIST \citep{JC2016}. These models take different approaches to modeling the 
complex effects of rotation, however they all rely on some common 
theoretical principles. Rotation is thought to alter the chemical mixing 
within stars due to several induced hydrodynamical instabilities (see e.g. 
\citealt{HL2000,MaMe2000b}). As a consequence, the core of a rotating star 
may gain access to a greater fuel supply than otherwise, leading to both 
greater luminosity and an extended lifetime. Furthermore, a rotating star 
may become significantly oblate due to latitudinally dependent centrifugal 
forces. As described by \cite{HvZ1924}, this effect, at times known as 
\emph{gravity darkening}, introduces a substantial viewing angle dependence 
to the observed color and magnitude of a star. Combined, rotationally 
enhanced chemical mixing and gravity darkening are able to substantially 
alter the color and magnitude position of key stellar population features 
on a CMD, particularly the main-sequence turn off (MSTO).

Classically, non-rotating isochrones and stellar models have been used to 
determine the ages of these benchmark clusters (e.g. \citealt{JM1981, MP1988, 
GM1993, MAP1998, JKMT2004}). However, recent studies have highlighted that a 
significant degree of uncertainty remains in how we model these systems. For 
instance, \cite{BH2015a} in fitting these clusters with a coarse grid of 
rotating Geneva stellar models, interpolated with a finer grid of
non-rotating PARSEC models (\citealt{LG2002}), discovered that the Praesepe
and Hyades may be older than previously thought. The effects of rotation, in
their modeling resulted in a best-fit age of $\sim800$ Myr, compared to the
classically inferred ages of $\sim600$ Myr found via non-rotating models.
This discrepancy motivates us to investigate the extent to which stellar
rotation affects key cluster parameters, such as age and metallicity, and
how extensively this uncertainty might exist across our stellar models.

The importance of rotation for interpreting open cluster CMDs extends well 
beyond these three benchmark clusters. Recent studies exploring the effects 
of rotation have been motivated by the potential for rotational effects to 
explain the extended main sequence turn offs (eMSTOs) of clusters residing 
in the Large and Small Magellanic Cloud (LMC and SMC); e.g. \cite{LG2013, 
MC2015}. It is still an ongoing debate as to the level that stellar rotation 
is responsible  for this phenomenon, e.g. \cite{BdM2009, PG2014, BH2015c, 
PC2017}. Certainly, as demonstrated by \cite{MaMe2000b, HL2000} and others, 
rotation can have strong effects on stars during and near the MSTO phase. 
The consequent flux and temperature alterations that result from rotational,
effects may cause an observer to perceive a MSTO that is collectively 
brighter or fainter in reality, compared to its theoretically non-rotating 
model. A brighter or younger MSTO mimics either a younger or older stellar 
population respectively. Furthermore, with a distribution of stars at 
various rotation rates and viewing angles, there is a possibility for the 
appearance of an eMSTO, as now a distribution of fainter and brighter stars 
(i.e., due to rotation) coexist within the population.

Furthermore, rotation can affect the integrated light of stellar populations 
as well. The impact of stellar rotation on bolometric luminosity and the 
ionizing spectra of massive stellar populations was explored by e.g. 
\cite{EML2012, CCB2017}. Both groups modeled the interplay between stellar 
rotation and the integrated light of galaxies whose spectral energy is 
dominated by the output from massive stars. In either case, rotation could 
enhance the ionizing radiation output of massive stars in quantity and 
duration (although to varying degrees, dependent on model assumptions). In 
affecting the population's composite spectral properties, which are tied to 
its inferred stellar mass and star formation history (see e.g., 
\citealt{CC2013} for a review), rotation may have far-reaching implications in 
extragalactic astronomy.

Here we present results derived from a self-consistent set of stellar 
evolution models, similar to the MIST models developed by \cite{JC2016} 
(see also \citealt{ADo2016}), but with a larger range in rotation rate as 
well a custom mass and metallicity range. These models require neither major 
interpolations nor extrapolations over stellar mass or metallicity, as has 
been required in the past. Following an overview of the data featured in 
\S \ref{s:data}, this paper presents a base description of the physics 
underlying our models (\S \ref{ssec:models}), leaving further details to 
the aforementioned MIST papers. We detail the methods used in applying our 
models to observed data through a statistical analysis package known as 
MATCH, written by \cite{AD2001}, in \S \ref{ssec:cmdfitting}. Subsequently 
we discuss the results of applying our models to observations of the Hyades, 
Praesepe, and Pleiades clusters in \S \ref{s:results}, and cast the 
implications of our findings in a broader context in \S \ref{s:discussion}. 
Finally, our work is summarized in \S \ref{s:summary}.

\section{Data}
\label{s:data}

In this section we provide a summary of the photometric data used in our CMD 
fitting for the Hyades (\S \ref{ssec:hyades}), Praesepe (\S
\ref{ssec:praesepe}), and Pleiades (\S \ref{ssec:pleiades}) clusters. We
also provide a brief summary of the salient properties of each cluster.
Although metallicities are cited, we do not adopt any of these listed values.
We only use them to compare with our CMD-derived metallicities in later
sections.

\begin{deluxetable*}{ccccc}[!ht]
\tablecaption{Adopted Cluster Parameters}
\tablehead{ \colhead{Cluster} & \colhead{$\mu$\tablenotemark{1}}
          & \colhead{$A_{\rm{V}}$\tablenotemark{2}} & \colhead{$B_T, V_T$ Binary Fraction}
          & \colhead{$J, K_s$ Binary Fraction \tablenotemark{3}}}

\startdata
  The Hyades & 3.349 & 0.0031 & 0.0 & 0.25 \\
   & \citep{GAIA2017} & \citep{BT2006} \\
  The Praesepe & 6.26 & 0.0837 & 0.0 & 0.30 \\
   & \citep{AG2009} & \citep{BT2006} \\
  The Pleiades & 5.64 & 0.1054 & 0.0 & 0.0\\
   & \citep{GAIA2016} & \citep{BT2008}
  \enddata
  \vspace{0.1cm}
  \tablenotetext{1}{Distance modulus}
  \tablenotetext{2}{Interstellar reddening}
  \tablenotetext{3}{NIR data has not been cleaned of binaries for the 
                    Hyades/Praesepe.}
\label{t:muav}
\end{deluxetable*}

\subsection{The Hyades}
\label{ssec:hyades}
Of our target clusters, the Hyades is nearest to our solar system, making it
a popular object of study amongst astronomers for many years (e.g.,
\citealt{WMS1939,PAW1967,vALH1997,SR2018}). We adopt a distance modulus
$\mu=3.349$ mag ($46.75\pm0.46$ pc) from \cite{GAIA2017} and an extinction
of $A_{\rm{V}}=0.0031$ \citep{BT2006}. Historically, the age of the cluster
has been determined to be around 600 Myr, according to non-rotating
isochrone fits; for example, \cite{MAP1998} derive an age of $625\pm50$ Myr
from fitting optical CMD data. \cite{ELM2018} measured Li surface abundance
in 6 brown dwarf candidates of the Hyades, and using the models of
\cite{IB2015}, derived an age of $650\pm70$ Myr. The Hyades may have a
[Fe/H]$=0.103\pm0.008$ according to \cite{TJ2005}; recently, \cite{JDC2017}
found $0.146\pm0.004$ from spectra of 37 Hyads.

Our optical data is comprised of the ``high fidelity'' stellar members
identified by \cite{deB2001}; these are stars with relatively high 
membership likelihood and evidence supporting their status as single stars. 
Binary systems are removed from this sample, and so our assumed binary 
fraction is zero. This catalogue is based on the main Hipparcos catalogue 
\citep{MAP1997}, utilizing derivations of secular parallaxes for cluster 
candidates to determine membership likelihoods. Using these high fidelity 
members, we fit stars with $V_T < 8$ magnitudes, forcing the fitting 
algorithm to focus on the age-sensitive MSTO.

Due to the close proximity of the Hyades  a significant spread in the 
apparent magnitudes will exist due to differential distance effects. This 
effect must be accounted for when fitting observed CMDs of the Hyades to 
models. \cite{deB2001} manage this by considering the absolute magnitude of 
stars (in their case derived using e.g. the Hipparcos secular parallaxes 
from the main catalogue, and recorded in their data tables). We do the same, 
but convert back to an apparent magnitude using the average distance modulus, 
$\mu$, of the Hyades cluster, $\mu=3.349$ mag. In essence, we place all 
stars at the mean distance of the cluster.

Our infrared data is sourced by the members of \cite{BG2013}, who used the 
members of \cite{SR2011}, where the former researchers made efforts to 
extend membership down to $0.1 M_{\odot}$. In both instances, membership was 
determined by the convergent point method. Furthermore, field star 
contamination has been estimated for these data sets; \cite{BG2013} found 
that contamination is likely, although decreasing to negligible levels 
(less than $~10\%$) inwards of $18$ pc within the catalogue. Binaries are 
present as well, and here we assume a binary fraction of 25\%, as suggested 
by \cite{JEG1988} (this value was also adopted by \citealt{SR2011}).

\subsection{The Praesepe}
\label{ssec:praesepe}
The Praesepe is the furthest of our target clusters at roughly four times 
the distance to the Hyades, but also appears similar to the Hyades in both 
its age and chemical composition. We adopt $\mu=6.26$ mag ($\approx179$ pc) 
\citep{AG2009} and $A_{\rm{V}}=0.0837$ \citep{BT2006} for modeling Praesepe. 
The metallicity of Praesepe has been estimated e.g. by \cite{AB2013} to be 
[Fe/H]$=0.12\pm0.04$ based on measurements of $11$ Praesepe dwarfs; 
\cite{JDC2017} found $0.156\pm0.004$ from spectra of 39 Praesepe members.

Our optical data derives from the 24 members tabulated by \cite{MDL2002}. 
We have cross matched these stars to obtain updated Tycho $B_T$, $V_T$ 
magnitudes via their Hipparcos ID numbers. This has been checked and cleaned 
of binary and field stars by the catalog's authors. Here we also impose a 
$V_T < 9$ magnitude cut, for the same reasons that we made this cut in our 
data of the Hyades (see \ref{ssec:hyades}).

Additionally, we use the near infrared (NIR) 2MASS $J$, $K_s$ magnitudes 
catalogued by \cite{PFW2014}. This data set includes 1040 stellar members 
in total, ranging from $M\approx 0.11-2.4 M_{\odot}$ acquired via proper 
motion analysis. A binary fraction of 20-40\% is suggested for these members; 
correspondingly, we adopt a binary fraction of 30\%. Furthermore, as noted 
by \cite{PFW2014}, Praesepe members exhibit distinct proper motions from 
potentially interfering field stars, virtually eliminating confusion between 
them. The authors estimate that their member list possesses $\sim16\%$ 
non-members (i.e., 862 true members and 168 non-members). Here we also 
impose a cut to only include stars with $V_T < 9$; focusing on the MSTO, 
however also lessening the probability of contaminating stars, as 
contamination is estimated to lessen towards the population's bright end.

\subsection{The Pleiades}
\label{ssec:pleiades}
The Pleiades is closer to us than the Praesepe, at roughly 3 times the 
distance to the Hyades, possessing an appreciably younger age, as well as a 
lower metallicity than the previous clusters. Our adopted distance modulus 
and extinction for this cluster are: $5.64$ mag \citep{GAIA2016}, or 
$\approx134$ pc, and $A_{\rm{V}}=0.1054$ \citep{BT2008}, respectively. 
\cite{DS2009} found [Fe/H]$=0.03\pm0.05$ from spectroscopic measurements of 
20 Pleiads. The age has been estimated by \cite{ByN2004} to be $130\pm20$ 
Myr using the LDB. From CMD analysis, \cite{GM1993} determined $100$ Myr by 
fitting non-rotating isochrones to its higher mass stars. Recently, 
\cite{SED2015} found an age of $112\pm5$ Myr via the LDB, using the 
evolutionary models of \cite{IB1998, IB2015}.

Our optical data is again found using the Hipparcos IDs of cluster members 
reported in \cite{MDL2002}, subsequently cross matched for $B_T$, $V_T$ 
magnitudes. Binary and field stars have been removed from this sample by 
\cite{LL2000} with a maximum likelihood estimation to determine likely 
(single) cluster members.

We also use the 2MASS magnitudes of members reported by \cite{JRS2007}. In 
fitting this data, we only include stars with $K_s < 5.0$ mag, for the same 
reasons that we made similar cuts in the Praesepe and Hyades (see 
\ref{ssec:hyades}). There is no estimate of binary fraction for this data 
but stars in this magnitude range are well known, higher mass members of the 
Pleiades. They are: Alcyone, Electra, Atlas, Maia, Merope, Taygeta, Pleione, 
and Celeno -- many of these are Be stars. Of the stars listed, Atlas is a 
double line spectroscopic binary system \citep{NZ2004}; evidence from lunar
occultation \citep{RCL1994} suggests Taygeta may be a binary system; Pleione 
may be a single line spectroscopic binary (see e.g. \citealt{JIK1996} and 
\citealt{JN2010}); Electra's multiplicity is inconclusive thus far, with 
contradicting evidence (e.g. \citealt{HAA1965} and \citealt{JP1971}); 
meanwhile Alcyone is a multiple star system, although most of its members 
are further than 77 arcsec (as listed in the Washington Double Star Catalog). 
In practice, we assume Atlas and Taygeta are systems whose photometry may be 
significantly affected by their companions and remove them. We make cuts in 
our optical data similar to the infrared data, only using stars with 
$V_T < 5.0$.

\begin{deluxetable}{lcc}[t!]
\tablecaption{Model Mass, [Fe/H], and Rotation Rate Range}
\tablehead{ \colhead{Mass Range} & \colhead{[Fe/H] Range}
          & \colhead{$\frac{\Omega}{\Omega_{\rm{c}}}$ Range}
          }

\startdata
  0.1 - 8.0 $M_{\odot}$ & -0.75, -0.60, ..., 0.45 & 0.0, 0.1, ..., 0.6 \\
\enddata
\vspace{0.1cm}
\label{t:modp}
\end{deluxetable} 

\section{Methodology}
\label{sec:methods}
Here we give an overview of our models and the fitting procedure from which 
we derive population age and metallicity. In \S \ref{ssec:models}, a brief 
overview of the background physics in our stellar models is given and the
effects of rotation and gravity darkening are discussed. Following this, \S
\ref{ssec:matchmod} introduces MATCH \citep{AD2001}, which we used to turn
our stellar models into composite stellar populations. Our fitting procedure
is also done via MATCH, fitting composite stellar populations to data, with
an outline of its fitting procedure given in \S \ref{ssec:cmdfitting}, along
with mock tests performed to demonstrate its accuracy.

\subsection{MESA Stellar Models}
\label{ssec:models}
Our models are fundamentally based on the MESA stellar evolution code
\citep{BP2011,BP2013, BP2015, BP2018}. MESA is a 1D, open source,
parallelized code, written with a modular design for flexibility in
customizing its incorporated physics. The stellar structure and composition
equations are simultaneously solved via a Newton-Raphson solver. Boundary
conditions are necessary for solving stellar structure equations; MESA
provides a number of options for their choice. In these models, the ATLAS12
code (\citealt{RLK1970,RLK1993}) model atmosphere tables set these boundary 
conditions; as done in \cite{JC2016}. Furthermore, we use the protosolar 
abundances of \cite{MA2009} where metallicities are scaled to 
$Z=Z_{\odot,protosolar}=0.0142$. In \S \ref{sssec:rotation}, we give an overview
of how rotation manipulates a star and \S \ref{sssec:gdark} is dedicated to
a description of gravity darkening and how it is implemented in our stellar
models.

In this work, we have expanded upon the grids presented in \cite{JC2016} by 
computing a set of models with fine variation in initial rotation rates and 
a denser grid in metallicity. These are not fully evolved, but truncated to 
the end of core helium burning (CHeB), with a metallicity range more focused 
on the solar neighborhood and LMC. Our metallicity range has a higher 
resolution around values appropriate for our clusters of interest (near 
solar [Fe/H] $= 0.0$). These choices were made to introduce greater 
variability in rotation rate as a model parameter while maintaining a 
reasonable model creation timescale. The mass, [Fe/H], and rotation ranges 
covered by our models are listed in Table \ref{t:modp}.

\begin{figure*}[!htb]
\center
\includegraphics[width=0.95\linewidth]{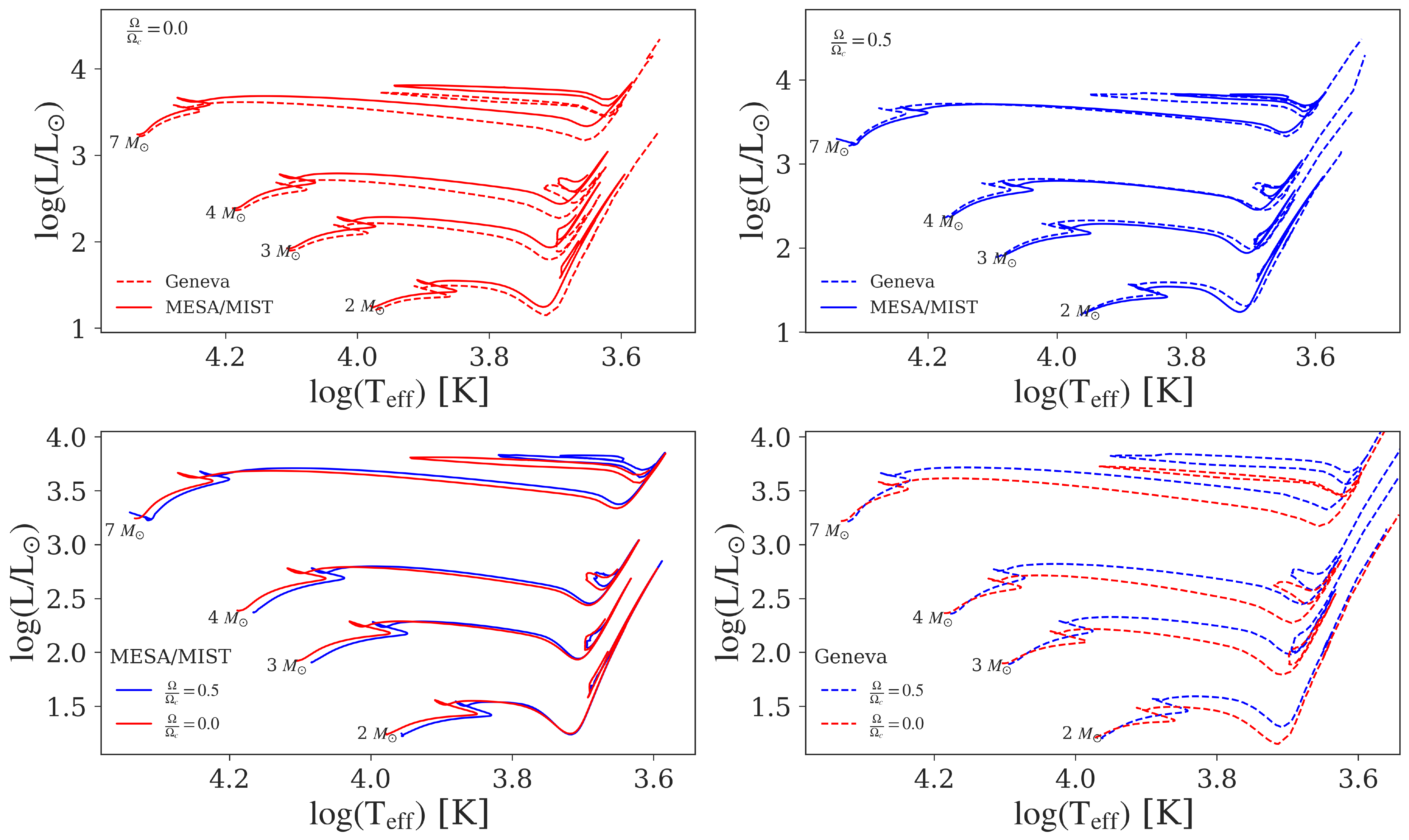}
\vspace{0.1cm}
\caption{Evolutionary tracks of our stellar models in comparison to 
  corresponding Geneva models, displaying model differences between the two 
  codes. The solid red (blue) lines are MIST non-rotating (rotating) models, 
  while dashed blue (orange)  lines are Geneva models. Rotating MIST models 
  tend to be fainter and cooler than their Geneva counterparts. (All tracks 
  have $i=0^{\circ}$, i.e., viewing at the equator.)}
\label{fig:trkcmp}
\end{figure*}

\subsubsection{Stellar Rotation}
\label{sssec:rotation}
In this section we provide a brief overview of the key physical processes 
related to modeling stellar rotation. See \cite{JC2016} and the MESA papers 
\citep{BP2011,BP2013,BP2015,BP2018} for details. In MESA, surface magnetic effects 
are not modeled, thus magnetic braking is neglected presently. To compensate 
and reproduce the observed slow rotation of solar-like stars, models below 
$1.2 M_{\odot}$ do not rotate. Between $1.2$ and $1.8 M_{\odot}$, rotation is 
scaled by a factor ranging from 0 to 1 as mass increases, such that models 
above $1.8 M_{\odot}$ rotate at their full velocity. For reference: MSTO 
masses range from $\sim1.4-2.5 M_{\odot}$ at isochrone ages of $650-750$ Myr 
(e.g. the Praesepe and Hyades), and from $\sim2.4-4.5 M_{\odot}$ at ages of 
$100-150$ Myr (e.g. the Pleiades), in our modeling.


Rotational velocity is commonly characterized by the ratio 
$v_{\rm{ZAMS}}/v_{c}=\Omega/\Omega_{c}$. This is the zero age main sequence 
(ZAMS) surface equatorial rotation rate, $v_{\rm{ZAMS}}$, compared with the 
the critical rotation velocity, $v_{c}$, the velocity that would  disrupt 
the star through centrifugal force. The quantities $\Omega$ and $\Omega_{c}$ 
are the angular counterparts.

The 1D code, MESA, implements rotation through the shellular approximation 
described by \cite{KT1970}, and employs the diffusive approximation 
introduced by \cite{ES1978} to model rotationally enhanced chemical mixing. 
The latter scheme requires that a choice be made for two parameters: $f_c$, 
which dictates how closely compositional mixing follows the flow of angular 
momentum transport, and $f_{\mu}$ which encodes the efficiency of rotational 
mixing in the presence of stabilizing molecular weight gradients. Values of 
$f_c=1/30$ (calibrated to reproduce the surface $^7$Li abundance of the Sun 
by \citealt{MHP1989,CZ1992}) and $f_{\mu}=0.05$ (found by \citealt{HLW2000} to 
reproduce nitrogen surface enhancement in evolved 10-20 $M_{\odot}$ stars 
from \citealt{GL1992,AH1993,MV2000}) are adopted in our models.

Although the shellular approximation is virtually ubiquitous in 1D stellar 
evolution codes, the diffusive approximation is not. Other codes, such as 
Geneva \citep{SE2012} take a diffusive-advective approach (as described in 
\citealt{MaMe2000a}) instead. Consequently, MESA models can exhibit 
noticeable distinctions from models developed via alternative codes, in 
some respects (as pointed out by \citealt{JC2016}, primarily showcasing the 
differences of higher mass models). Rotating MIST models tend to be fainter 
and cooler at the MSTO compared to those same models computed by Geneva, 
perhaps due to less efficient chemical mixing from rotation effects.

In Figure \ref{fig:trkcmp} we reiterate these points, focusing on lower and
intermediate masses, as would likely be found in our relatively young target
clusters. In these plots we show evolutionary tracks for stellar models of
$2, 3, 4$ and $7 M_{\odot}$. Our models are displayed as solid lines, while 
dashed lines correspond to Geneva models. Rotation rate is encoded by color, 
wherein red represents $\Omega/\Omega_{\rm{c}}=0.0$ (non-rotating) and blue is 
$\Omega/\Omega_{\rm{c}}=0.5$. The two top panels overlay our models and Geneva
models with no rotation (top left) and with rotation (top right); these top 
panels are comparisons across model sets. Our non-rotating models are 
brighter and hotter than non-rotating Geneva models, however the roles are 
reversed in comparing rotating models. The two bottom panels compare 
non-rotating to rotating models within each model set to see how rotation 
affects the models differently according to each respective code. Our models 
become cooler as the rotation rate increases (due to gravity darkening; \S 
\ref{sssec:gdark}), whereas Geneva models become hotter and brighter as 
their rotation rate is increased.

\begin{figure*}[!htb]
  \center
    \includegraphics[width=0.95\linewidth]{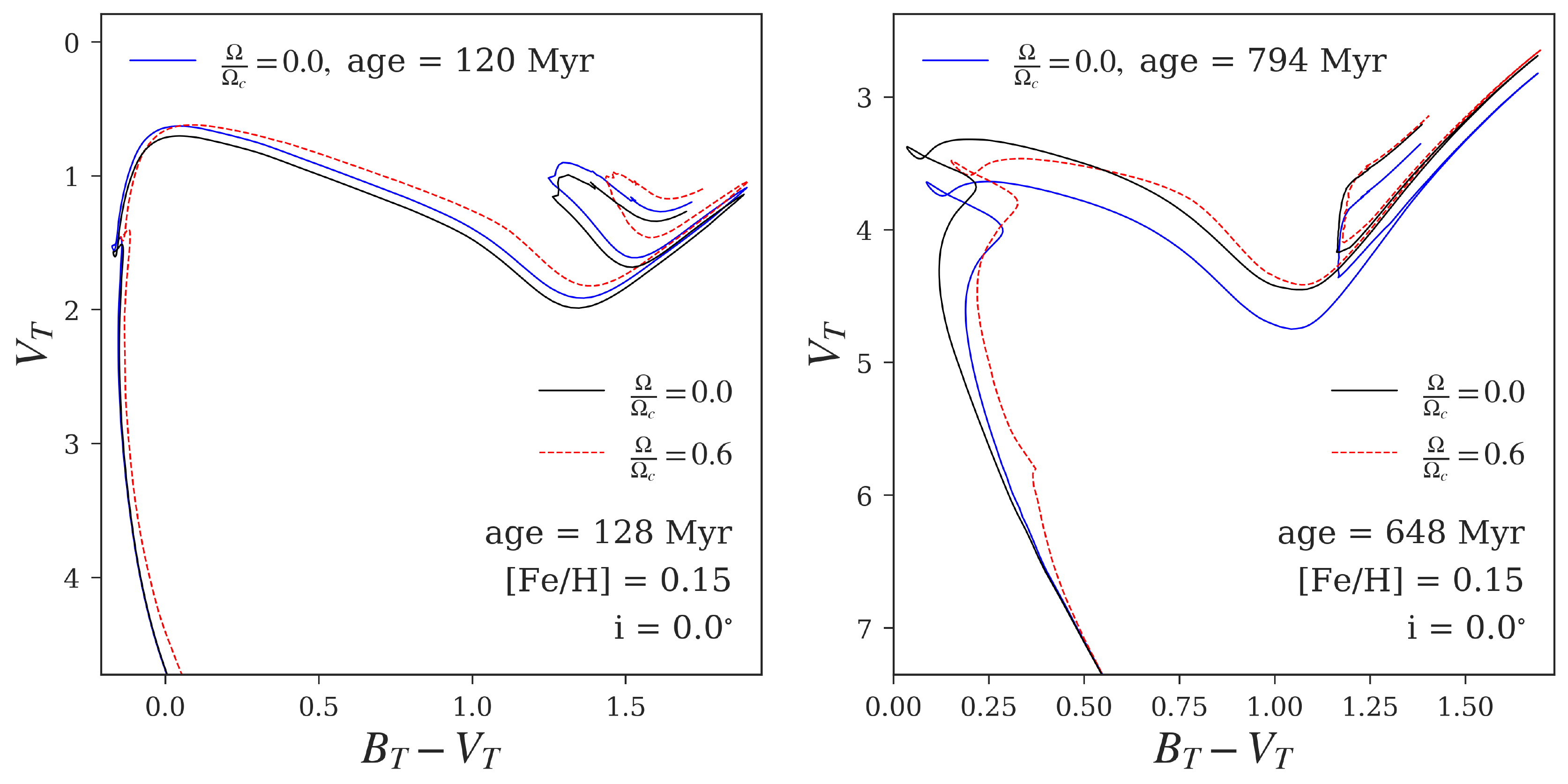}
    \caption{Displaying the effects of increasing $\Omega/\Omega_{\rm{c}}$ 
      from $0.0$ to $0.6$ on a [Fe/H] $= 0.15$ isochrone. The case of a 
      younger age near what is suggested for the Pleiades is shown at left, 
      the case of an older age near classical suggestions for the Hyades 
      and Praesepe at right. In each panel, a non-rotating isochrone is 
      displayed in solid black to compare with a rotating model 
      at $\Omega/\Omega_{\rm{c}}=0.6$ (red, dashed). In the left panel, increasing 
      $\Omega/\Omega_{\rm{c}}$ from 0.0 to 0.6 causes the non-rotating population 
      to move towards resembling a younger population (blue, solid). In the 
      right panel, increasing $\Omega/\Omega_{\rm{c}}$ creates a cooler MSTO, 
      similar in color to an older, non-rotating population (blue, solid); 
      although the rotating population remains brighter than this older, 
      non-rotating population. (All models have $i=0^{\circ}$ here, i.e., viewing 
      at the equator.)}
\label{fig:vvcvary}
\end{figure*}

\begin{figure*}[!htb]
\center
\includegraphics[width=0.98\linewidth]{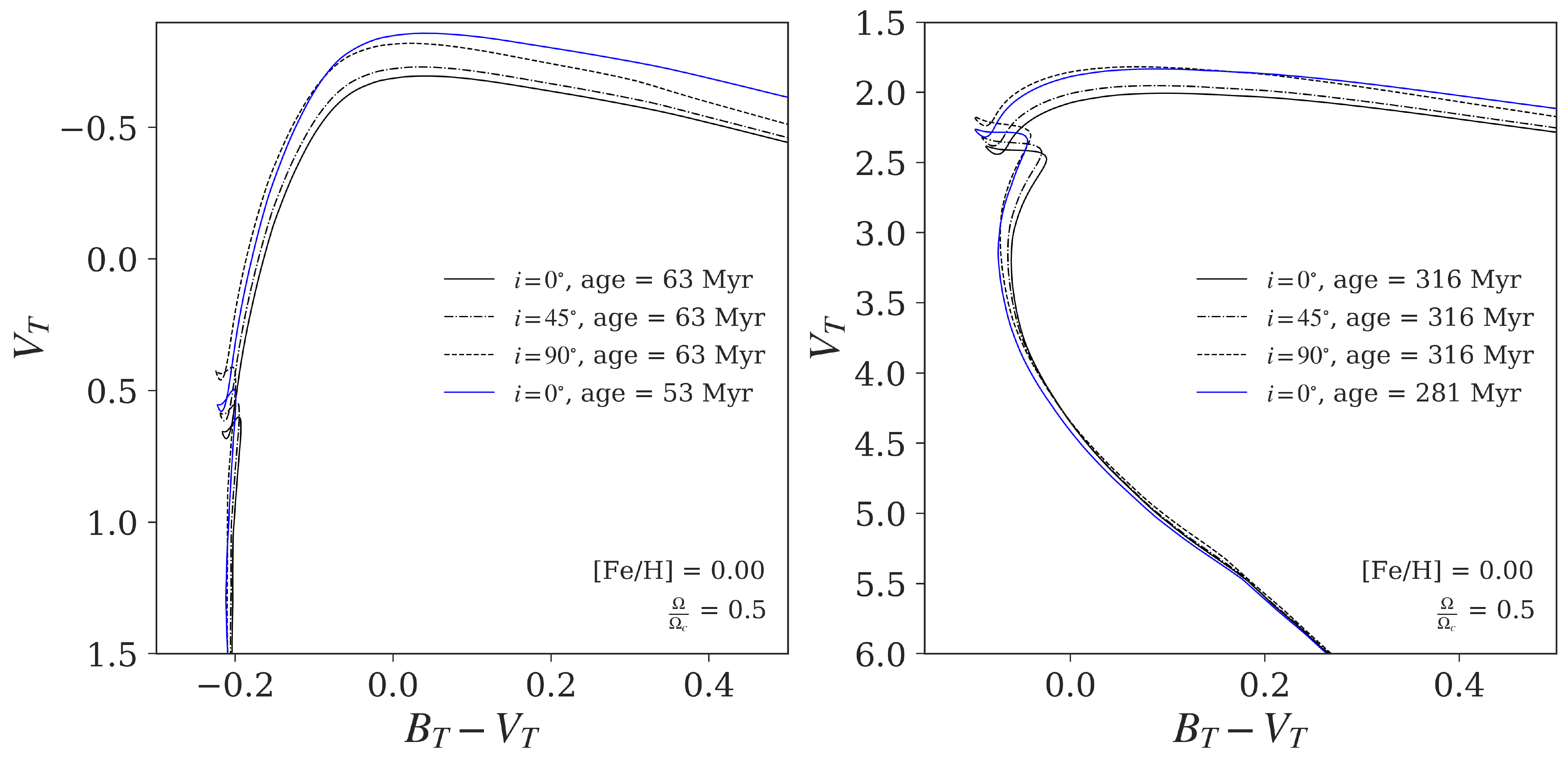}
\vspace{0.1cm}
\caption{Demonstration of the effects of gravity darkening on a CMD, meant
  to demonstrate general trends of the effects. Gravity darkening is viewing
  angle dependent and can issue an alteration of several mmag in color and
  ~0.2 mag in brightness. The cases for viewing angle $i = 0^{\circ}$
  (equator, solid), $i = 45^{\circ}$ (dash-dot), and
  $i = 90^{\circ}$ (pole, dashed) are shown. In affecting both the 
  luminosity and temperature of stars, these effects can make MSTO stars 
  appear younger than they intrinsically are. A slightly younger isochrone 
  at $i = 0^{\circ}$  is shown (blue, solid) for comparison.}
\label{fig:gdark}
\end{figure*}

Figure \ref{fig:vvcvary} shows these effects on a CMD, as they manifest in 
the MSTO of an isochrone constructed from our models. Figure 
\ref{fig:vvcvary} shows a $\Omega/\Omega_{\rm{c}}=0.0$  isochrone (black) compared 
to a $\Omega/\Omega_{\rm{c}}=0.6$ isochrone (red, dashed) at an age near what has 
been classically reported for the Pleiades, $\sim100$ Myr, (left panel) 
and an age near classical reports for the Hyades/Praesepe, $\sim600$ Myr 
(right panel). The effects are visually quite different between these 
ages. On the left, a slightly younger isochrone is shown in solid blue to 
point out how rotation can enhance the MSTO brightness, causing an older 
rotating population to mimic the morphology of a younger non-rotating one; 
here our rotating models might predict older ages compared to non-rotating
models (albeit, the effect is very slight in our modeling). Whereas on the right, 
less massive and more slowly rotating stars exist on the MSTO at $600$ Myr and the 
effects of rotation are modest. Thus, the MSTO barely becomes brighter at these 
older ages, and primarily appears cooler, mimicking the color-position of an older 
population (shown as blue in the righthand panel); here our models might find a 
younger age (decreasing age will make the MSTO hotter again) than predicted by 
non-rotating models, or could cause shifts in derived metallicity. Hence, our 
models do not behave as straightforwardly as Geneva models, where rotation 
primarily makes the MSTO brighter and hotter. The behavior exhibited by Geneva 
models leads to rotating models predicting younger ages than non-rotating models.

At $100$ Myr, MSTO stellar masses are $\sim2.4-4.5 M_{\odot}$, which rotate 
fully. Meanwhile at $600$ Myr, the extant MSTO mass range is roughly 
$\sim1.4-2.5 M_{\odot}$. As discussed at the start of this section, stars 
below $1.8 M_{\odot}$ do not rotate at full capacity in order to replicate 
the effects of magnetic braking relevant to lower mass stars with 
convective envelopes. Thus, the effects of rotation on MSTO morphology do 
change with age for our models.

That rotating MIST models are cooler and less luminous than their Geneva
counterparts (top right panel, Figure \ref{fig:trkcmp}) may stem from 
divergent approaches to rotationally induced angular momentum transport 
(such differences were also pointed out by \citealt{JC2016}). Meanwhile, 
non-rotating MIST models appear more luminous than those of Geneva (top 
left panel, Figure \ref{fig:trkcmp}), perhaps due to differing assumptions 
made for the treatment of convection.

Convective core overshoot (CCO) mixing is a chief aspect of convection
that can influence the MSTO's CMD morphology and position. CCO is the idea
that convectively driven material should not suddenly stop at the
theoretical boundaries (e.g., Schwarzschild or Ledoux) of a convective zone.
Rather, it seems feasible that momentum should carry material past these
boundaries, allowing it to penetrate into non-convective zones, see
e.g., \cite{KHB1963}. The penetration distance, often considered as an 
extension of the convective zone, is often either a fraction of the 
pressure scale height (denoted $H_{\rm{P}}$), or is modeled as a diffusive 
process with an exponentially decaying diffusion coefficient. We adopt the 
latter formalism, while Geneva adopts the former. Our adoptions (see \S 
3.6.1 of \citealt{JC2016}) are roughly equivalent to an extended convective 
core boundary of $0.2 H_{\rm{P}}$ (\S 2.1 of \citealt{ZM2010}), effectively twice 
the value adopted in Geneva.

Similar to rotationally enhanced mixing, CCO mixing can supply the stellar 
core with more fuel, enhancing energy production and the MS lifetimes of 
stars (e.g., \citealt{AM1975}), thereby affecting a population's MSTO on a 
CMD. A studies presented by \cite{CG2003,JHW2003} and \cite{GB2003} looked 
at the intermediate age clusters NGC 2173, SL 556, and NGC 2155, with 
non-rotating Padova (\citealt{LG2000}) and Yonsei-Yale (based on tracks 
from YREC, \citealt{DBG1992}) stellar models. Modifying the efficiency of 
CCO appeared to provide one avenue for the models to explain the observed 
CMD features. Although, an alternative perspective on CCO is that material 
outside the convective core (CC) boundary, once mixed via overshoot or some 
other process, may drive an increase in opacity or erasure of composition 
gradients, subsequently turning sub-adiabatic material super-adiabatic, and 
expanding the CC (discussed further in \S 2 of \citealt{BP2018}). CCO may 
partly be the result of a poorly defined CC boundary in our models. The 
proper treatment of CCO (and convection in general) is another active 
branch of development in stellar modeling. 

Of course, CCO is not a rotational effect, but is discussed here to expound
on the behavior shown in Figure \ref{fig:trkcmp} and to highlight its 
important (albeit uncertain) role in stellar evolution. Our models possess a
stronger mixing due to CCO (see \citealt{JC2016}, \S 3.6.2, and
\citealt{SE2012}, \S 2.3). As CCO mixing occurs regardless of rotation, this
is a possible cause for our non-rotating models maintaining a higher
luminosity than their non-rotating Geneva counterparts.

In addition to the differences shown here on a Hertzsprung-Russell diagram 
(HRD) and CMD, there are also differences in the MS lifetime extension 
afforded by enhanced rotational mixing. These differences were shown in 
Figure 20 of \cite{JC2016} where the ratio $\tau_{\rm{MS,R}}/\tau_{\rm{MS,NR}}$ (MS 
lifetime with rotation over the same without rotation) is plotted against 
the initial stellar mass $M_i$ (in units of $M_{\odot}$). Rotating Geneva 
models with $M_i > 2 M_{\odot}$ show a lifetime extension of roughly 
$\sim25\%$, while corresponding MIST models only garner an increase of
$\lessapprox10\%$.

The physical adoptions made in Geneva do not equate to those made in MIST. 
The uncertainties present in stellar modeling (convection and rotationally
enhanced mixing here; see, e.g., \citealt{SC2017} for a review) provide enough
leeway for inconsistent model behavior. As will be seen, our findings are
different from what was found in \cite{BH2015a,BH2015c}, likely due to model
differences. 

\subsubsection{Gravity Darkening}
\label{sssec:gdark}
In addition to enhanced chemical mixing, rotation also introduces oblateness 
to stellar structure. Latitudinally dependent centrifugal forces result in 
an oblate deformation of the star. The surface gravity, $g$, of a star is 
lessened by these forces leading to an effective, latitudinally dependent 
value, $g_{\rm{eff}}(\theta)$. Here, $\theta$ refers to the polar angle in a 
spherical coordinate system. The effective gravity of a rotating star is 
related to its radiative luminosity (e.g., as encapsulated by the von Zeipel 
theorem; \citealt{HvZ1924}), leading to a relation 
$g_{\rm{eff}} \propto T^{4}_{\rm{eff}}(\theta)$ between the effective gravity and 
temperature of a star. Hence, the observed colors (temperatures) and 
magnitudes (luminosities) become dependent on the viewing angle of 
observation. This effect is commonly termed gravity darkening.

The essential physical arguments of \cite{HvZ1924} qualitatively adhere well 
to observations, but the power law relation presented in the von Zeipel 
theorem may be too simplistic. As it was derived based on radiative flux 
relations, the von Zeipel theorem does not directly apply to stars with 
convective envelopes; \cite{LBL1967} derived a more general 
$g_{\rm{eff}} \propto T^{\beta}_{\rm{eff}}$, with $\beta=0.08$ for convective 
stars. More recently, in comparison to interferometric observations of 
rapidly rotating stars \citep{HAM2005, JA2006, GvB2006, MZ2009} it has been 
found that $g_{\rm{eff}} \propto T^{4}_{\rm{eff}}$ over estimates temperature 
variation going between pole and equator. In light of this, \cite{ELR2011} 
were motivated to derive a new formulation that can describe gravity 
darkening; they do so, deriving a relation that depends only the rotation 
rate of the star, at a given viewing angle, luminosity, and $T_{\rm{eff}}$.

We use the equations of \cite{ELR2011} to translate the model stellar 
luminosity $L$ and effective temperature $T_{\rm{eff}}$ across desired viewing 
angles. The symbol $i$ denotes the inclination (or viewing) angle in this 
paper; $i=0^{\circ}$ corresponds to an observation directed at the equator, 
whereas $i=90^{\circ}$ is directed at the star's pole.

A demonstration of gravity darkening's effects is shown in Figure 
\ref{fig:gdark} on our MESA models. Several isochrones are displayed for two 
scenarios: $\sim60$ Myr models are shown in the left panel and $\sim300$ Myr 
in the right. In both cases, black shows an older isochrone while blue marks 
a slightly younger isochrone. The solid black line shows $i=0^{\circ}$, dashed 
is $i=45^{\circ}$, and dot-dashed is $i=90^{\circ}$. In each case, the MSTO of 
an older isochrone moves towards mimicking that of a younger isochrone as 
the inclination angle varies from $0^{\circ}$ to $90^{\circ}$, due to the 
increased luminosity and temperature of the stellar pole. Conceivably, if 
stars were to host a distribution of various inclination angles, there is 
the possibility that these effects would create a broadened MSTO.

\subsection{MATCH Composite Populations}
\label{ssec:matchmod}
The final models that we fit to data are constructed in MATCH 
\citep{AD1997, AD2001}, a tool used to study resolved stellar populations 
(e.g. see \citealt{JdJ2008, DW2011, DG2011}). MATCH uses a given set of 
stellar models (our MESA models in this case) to create its own library of 
isochrones. These models have a finite resolution in age and metallicity; 
we have chosen 0.02 dex for both. Thus, it should be noted that there is an 
inherent spread to the models, as the isochrone parameters do not pertain to 
delta functions in MATCH. Our composite stellar populations are constructed
with the effects of gravity darkening via random viewing angles and we
have also developed the ability for our synthetic populations to possess a
distribution of rotation rates.

CMDs are populated according to an initial mass function (IMF), describing 
the occurrence of stellar masses, which are subsequently combined into a 
composite model population. For this purpose we specify a Kroupa IMF 
\citep{PK2001}. Additionally, MATCH is able to consider binary systems when 
drawing its models, given a binary fraction. Binaries are added 
according to a flat distribution in mass fraction (i.e., from 0 to 1). So 
long as both stars are alive in the binary, magnitudes are the sum of fluxes 
from each star; if the primary has died, the magnitude is of the secondary 
survivor only. Colors are computed by constructing the magnitude in each 
filter in this way, and then taking the difference of the magnitudes. MATCH 
does not model interactions between binary companions. In the following 
discussion, we will list our adopted values of binary fraction where 
appropriate (adopted binary fractions are also listed in Table \ref{t:muav}).

\begin{figure}[!htb]
  \centering
  \includegraphics[width=0.95\linewidth]{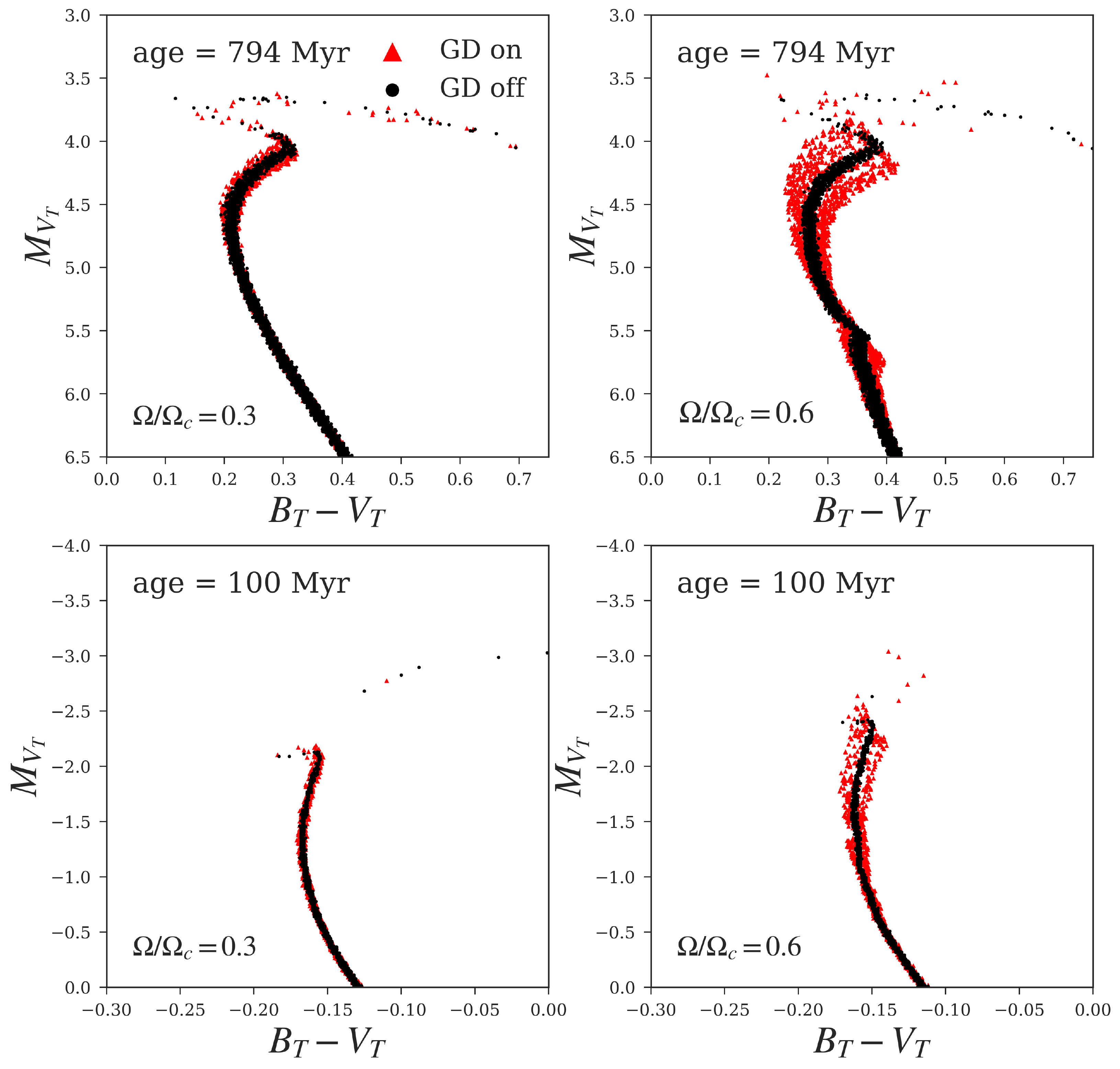}
  \vspace{0.1cm}
  \caption{Differences between the effects of gravity darkening (GD) at
    $\Omega/\Omega_{\rm{c}}=0.3$ (left column) and $\Omega/\Omega_{\rm{c}}=0.6$ (right 
    column). The effect is stronger for stars that rotate faster, leading to 
    a greater broadening of the MSTO as rotation rate increases. Age is 794 
    Myr in the top row, 100 Myr in the bottom row, and [Fe/H] = 0.15. These 
    simulated clusters have masses of roughly $1\times10^5 M_{\odot}$.}
\label{fig:modgd}
\end{figure}

\begin{figure*}[!htb]
  \centering
  \includegraphics[width=0.95\linewidth]{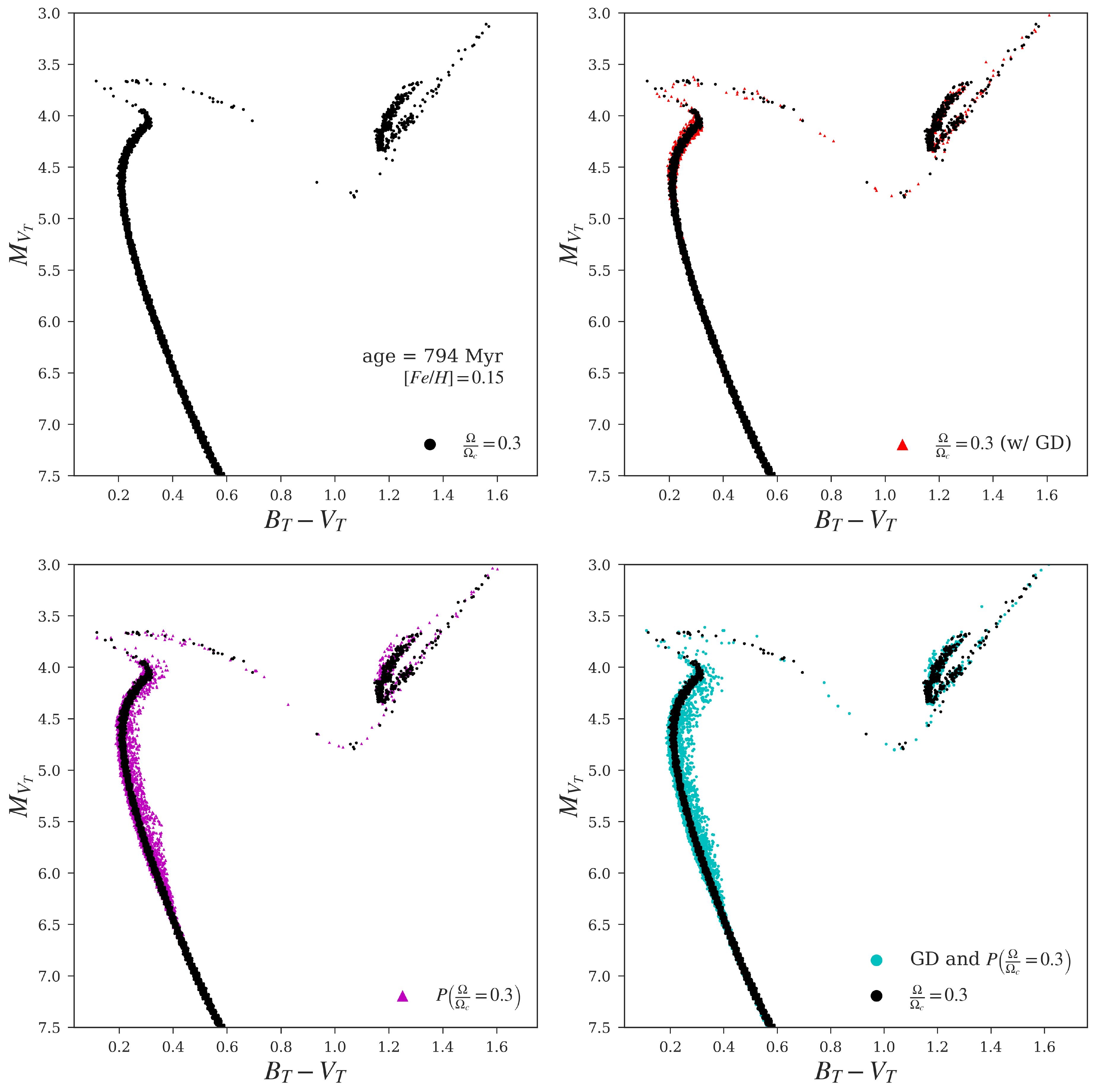}
  \vspace{0.1cm}
  \caption{Effect of rotation on simulated color-magnitude diagrams. The 
    effects due to including gravity darkening only (top right; GD), a 
    Gaussian distribution of rotation rates (bottom left), and finally the 
    inclusion of both phenomena (bottom right). This is all in comparison to 
    what a model population created through MATCH where none of these 
    effects are modeled looks like (top left), shown in black. All synthetic 
    populations shown correspond either to a singular $\Omega/\Omega_{\rm{c}}=0.3$ 
    for the population, or to a mean $\Omega/\Omega_{\rm{c}}=0.3$ in the case of 
    the rotation distribution (with standard deviation $0.2$ dex). These 
    models possess a finite age and metallicity resolution: 0.02 dex for 
    each, lending to an inherent spread in their morphologies.}
\label{fig:moddemo}
\end{figure*}

The ability to model gravity darkening was developed and added to MATCH for 
this project, where the brightness and temperature of our models are altered 
according to the equations of \cite{ELR2011}. These alterations are a 
function of the model's viewing angle (which is drawn randomly as it is 
added to the composite stellar population under construction) and their 
rotation rate. In order to demonstrate these effects, we have created 
artificial data sets using the MATCH program ``fake'', which generates a set 
of photometric data for use on a CMD, given some finite period of star 
formation and metallicity variation. This program is also able to simulate 
photometric errors, but we take the errors to be zero in all cases, in order 
to highlight rotation related effects. For this reason, we have also 
increased model resolution to 0.01 dex in Figures \ref{fig:modgd} and 
\ref{fig:moddemo}, to lessen the spread due to finite composite population 
model age and metallicity resolution and create an appearance closer to an 
SSP.

Effects from gravity darkening are shown in Figure \ref{fig:modgd}, as 
incorporated in MATCH. In Figure \ref{fig:modgd}, an artificial data set 
that neglects gravity darkening is shown in black, in comparison to an 
artificial data set that does model gravity darkening which is shown in red. 
Several scenarios are demonstrated: the left column shows isochrones with  
$\Omega/\Omega_{\rm{c}}=0.3$, while the right shows $\Omega/\Omega_{\rm{c}}=0.6$. 
Meanwhile, the top row compares the effects of gravity darkening at 794 Myr 
(nearer to the ages of the Hyades and Praesepe), and the bottom row shows 
100 Myr (near the age of the Pleiades). The inclusion of gravity darkening 
has a much stronger effect at higher rotation rates, issuing a greater 
spread to the MSTO.

Furthermore, stellar populations likely do not have stars rotating at a 
fixed value, rather they appear to possess a distribution of rates. 
Studies such as \cite{ZR2012} and \cite{RZG2007} have found evidence for a 
bimodal distribution of rotation rates for A and low mass B-type stars, with 
masses estimated around $1.5-3 M_{\odot}$. This mass range is roughly 
appropriate for the MSTO stars in our target clusters (mentioned at the 
beginning of \S \ref{sssec:rotation}). However, there is evidence that this 
distribution may change with stellar type, where e.g. B and O-type stars 
appear to exhibit a singly peaked asymmetric distribution (e.g. 
\citealt{P1996,HGMcS2010,RA2013}).

For this work, the ability to model a distribution of rotation rates has 
been incorporated as a new feature in MATCH. We use a Gaussian distribution 
as a preliminary choice to model these effects. As the rotation rates of 
our models range from $\Omega/\Omega_{\rm{c}}\in\left[0.0,0.6\right]$, we choose a
distribution with mean $0.3$ and standard deviation $0.2$. This distribution 
is truncated at the $\Omega/\Omega_{\rm{c}}$ parameter bounds of our model grid;
hereafter this distribution is referenced as 
$P\left(\Omega/\Omega_{\rm{c}}=0.3\right)$. The rotation rate of a proposed model 
is drawn from this Gaussian distribution as it is added to the composite 
population.

Although, as may be seen in Fig. 8 of \cite{HGMcS2010}, for the mass range 
$2.2 \leq M/M_{\odot} \leq 4.0$, stars may possess a distribution of 
$v/v_c$ that is peaked nearer to 0.6 (derived from $v\rm{sin}i$ measurements 
acquired from spectra). This mass range roughly corresponds to the MSTO of 
the Pleiades, for instance, and so our chosen distribution may not fully 
represent the distribution of rotation rates in this cluster. In order to 
test the effects of a distribution that includes higher rotation rates, we 
have also tested a flat distribution of $\Omega/\Omega_c$ between 0.0 and 
0.6. Given that our models are currently limited to $\Omega/\Omega_c = 0.6$, 
we opt for this rather than creating a new distribution centered on 
$\Omega/\Omega_c=0.6$. Generally, this flat distribution finds ages within 
10 Myr and metallicities very similar to those found with our Gaussian 
distribution. Our usage of a flat distribution is not an extensive test of 
a distribution including higher rotation rates, for instance, as may be seen 
in Fig. \ref{fig:modgd} the effects of rotation can vary dramatically towards 
higher rotation rates, so lack of very fast rotators may neglect a wide range 
of model behavior. However, usage of a flat distribution is intended to give 
some preliminary sense of whether results would change significantly in the 
presence of a greater number of fast rotators or not. In future work, we plan 
to include models from $\Omega/\Omega_c=0.0$ to $0.9$, allowing us to test a 
wider range of possible rotation distributions.

In Figure \ref{fig:moddemo}, four panels display a progression of the 
consequences from including gravity darkening (top right), a distribution 
of rotation rates (bottom left), and finally both together (bottom right), 
in comparison to what a synthetic population looks like in absence of these 
phenomena (black points). Models excluding a distribution of rotation rates 
are given the fixed value $\Omega/\Omega_{\rm{c}}=0.3$. With the relatively modest 
value of $\Omega/\Omega_{\rm{c}}=0.3$, gravity darkening has a weak influence on 
MSTO morphology, in comparison to that of a distribution of rotation rates 
(compare the top right and bottom left panels); although bear in mind that 
the effects of gravity darkening would increase with $\Omega/\Omega_{\rm{c}}$, 
as was shown in \ref{fig:modgd}.


\subsection{CMD Fitting}
\label{ssec:cmdfitting}
Here we present our CMD fitting methodology. First, we describe MATCH and 
give an overview of its operation in fitting stellar models to data (\S
\ref{sssec:match}). Next, we discuss mock tests that were performed to
demonstrate the accuracy of our results, using simulated observations that
have stellar densities similar to our target clusters (\S \ref{sssec:mock}).

\newpage

\subsubsection{MATCH}
\label{sssec:match}
CMD fitting is carried out via MATCH \citep{AD1997, AD2001}. This package 
includes the ability to fit for star formation histories and key population 
parameters (metallicity, distance, extinction). We exploit these 
capabilities to determine the age and metallicity of our target clusters, 
fixing distance and extinction to values based on existing literature. We do 
not solve for star formation history, instead running MATCH in ``ssp'' mode, 
solving for simple stellar populations (SSPs; populations assumed to form 
all stars in one burst of star formation, as is often assumed for open 
clusters).

With a given photometric data set (color and magnitude), and a collection 
of stellar models (MESA in our case), MATCH generates and compares Hess 
diagrams of the models and data. The model (i.e., synthetic stellar 
population constructed from supplied models) that best reproduces the 
observed stellar densities in the CMD space of the data is found with a 
Poisson likelihood statistic. This is calculated binwise in the Hess diagram, 
and combined to produce an overall likelihood for the model-data comparison.

Hess diagram bin size is specified by the user, and though guidelines exist,
this aspect of the fitting process ultimately involves a degree of personal
judgment. Generally, one wants to avoid choosing a bin size so large that 
the Hess diagram smooths out morphology in important features (like the 
MSTO), and not so small that spurious population features arise due to 
outliers, for example. We set a bin size of 0.10 in magnitude and 0.05 in 
color.

We derive two sets of ages and metallicities in this work. When deriving 
fits, we always include the effects of gravity darkening and rotationally 
enhanced chemical mixing. One set of results purely examines how the 
effects of stellar rotation on derived age and metallicity as 
rotation rate, $\Omega/\Omega_{\rm{c}}$ is varied. This is important since, as 
shown in Figure \ref{fig:modgd}, if the majority of stars rotate near 
$\Omega/\Omega_{\rm{c}}=0.6$, the effects of gravity darkening can be much 
greater. Our other set of results examines what happens when assuming a 
distribution of rotation rates. Our adopted distribution is a Gaussian, 
described in \S \ref{ssec:matchmod}.

In fitting populations at a fixed rotation rate, we fit seven populations 
constructed with a single value from $\Omega/\Omega_{\rm{c}}\in\left[0.0,0.6\right]$ 
(steps of 0.1 dex) separately. We select the population that produces the 
highest maximum posterior probability from these seven models as the 
best-fit. To examine the case where stars possess a Gaussian distribution of 
rotation rates (described in \S \ref{ssec:matchmod}), we simply take the set 
of best-fit age and metallicity that produces the highest posterior 
probability according to this model, providing a second set of derived age 
and metallicity.

\begin{figure*}[!htb]
  \centering
  \includegraphics[width=0.95\linewidth]{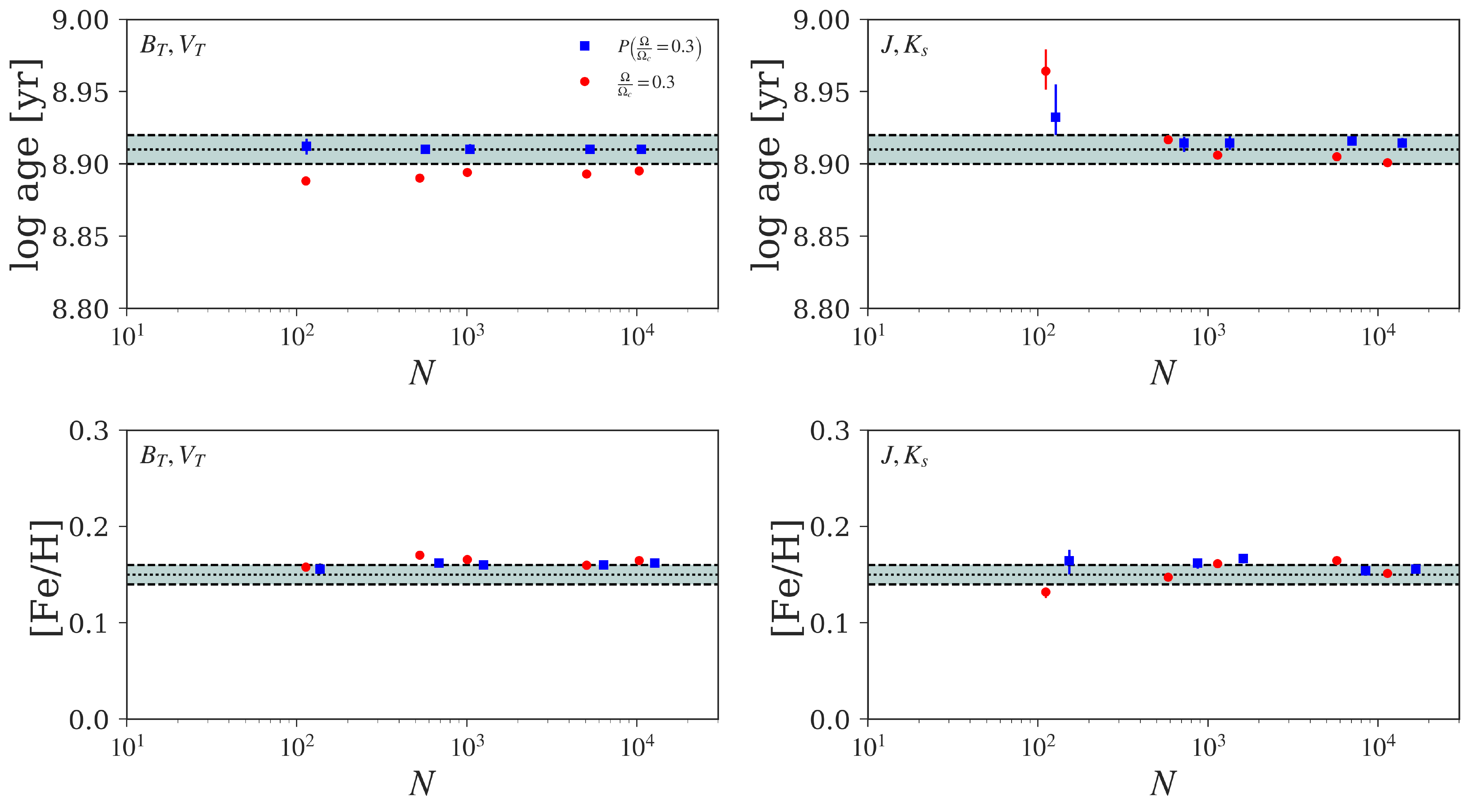}
  \vspace{0.1cm}
  \caption{Recovery of parameters from simulated CMDs. MATCH best-fit age vs. 
    number of artificial population members, N, in $B_T$, $V_T$ (left panels) 
    and $J$ $K_s$ (right panels). The blue region marks the truth, defined 
    as a burst of star formation taking place between $log$ age = 8.90 and 
    8.92. Likewise, the population metallicity is centered on 0.15 dex, with 
    a 0.02 dex spread in values. Each data point corresponds to an average 
    of 10 trial runs and errors are the average errors of the trial runs. 
    Dotted black lines indicate the middle of an age or metallicity bin.}
\label{fig:xvsn}
\end{figure*}

\subsubsection{Testing Parameter Recovery with Mock Data}
\label{sssec:mock}
Here we perform a number of mock tests in order to demonstrate the level of
accuracy that MATCH may achieve in fitting our models to data. A rotation
distribution and the modeling of gravity darkening are new additions to 
MATCH, and our data is fairly sparse in comparison to what MATCH is 
typically used for. Thus, we apply our models to an artificial data set 
(described below) and test for the ability of MATCH to recover the input age 
and [Fe/H] values used to construct this artificial data. We vary the number 
$N$ of artificial data points (i.e., $N$ is the number of stars in our 
artificial data) to test accuracy as the observations become more sparse.

An artificial data set with parameters similar to the Hyades, namely:
$\rm{log}$ age $= 8.90$, [Fe/H] $= 0.15\pm0.01$, $\mu = 3.34$,
$A_{\rm{V}}=0.0031$, was fit using our models in conjunction with MATCH. The
artificial data is created using the previously mentioned MATCH program 
``fake''. In these tests, our artificial clusters contain no multiplicity, 
and so our binary fraction is set to zero. Below are the results of several 
trial runs, fitting for the artificial cluster age and metallicity.

Figure \ref{fig:xvsn} shows the results of these mock tests. Here we plot 
the best-fit age (top row) and metallicity (bottom row) vs. N, where each 
point represents an average of 10 trial runs (to average over stochastic 
fluctuations). Red points correspond to models assigned a single rotation 
rate, fit to artificial data at a single rotation rate. Blue points 
correspond to models where a Gaussian distribution of rotation rates was 
used, fit to artificial data created with a Gaussian distribution of 
rotation rates. Errors shown are the average errors of the 10 trial runs. 
These errors were calculated via analysis of resultant posterior probability 
distribution, marginalized for the corresponding parameter; i.e., from 
the 16\% and 84\% percentiles of the posterior. The true values are 
represented by the light blue horizontal band, bounded by dashed black 
lines. For these tests, we are using the same models that will be used to 
fit the observations of our target clusters, which have a resolution 0.02 
dex in age and metallicity. We create stars in a single formation episode 
whose duration in time and spread in metallicity is 0.02 dex (matching
the resolution of our model composite stellar populations). Hence, the 
input age and metallicity bins, represented by the blue bands in Figure 
\ref{fig:xvsn}, span from log age = 8.90 to 8.92, and [Fe/H] = 0.14 to 0.16 
dex, respectively.

Ranging from artificial clusters comprised of several hundred to several 
thousand total stellar members, the input cluster parameters are often found, 
or are off by at most 0.05 dex or so. This range in stellar numbers is
chosen to replicate that of our target clusters. For instance, our sample of
the Hyades contains lists of roughly one to several hundred stars in optical
and NIR, respectively; the Pleiades lists contain several hundred; 
our list for the Praesepe consists of only 24 stars at the cluster turn off 
in the optical (although with $\sim$1000 total members in infrared for 
Praesepe). As seen in Figure \ref{fig:xvsn}, log age appears to not be 
recovered well in 2MASS $J$, $K_s$ for the case $N=100$ or so. However our 
NIR data lists contain nearer to $1000$ total members. 

Models become more degenerate in age and metallicity with fewer total stars, 
as both the MSTO and MS become less populous under IMF sampling. Accordingly, 
it is expected that the recovered parameter uncertainties should become 
larger as $N$ is decreased, as may be seen in Figure \ref{fig:xvsn}. Still, 
the error bars remain relatively small (of order $\lesssim0.03$ dex for log 
age and [Fe/H] for the worst cases) for results derived with the lowest 
stellar densities tested here.

\section{Results}
\label{s:results}
In this section we turn to modeling the benchmark open clusters: the Hyades,
Praesepe, and Pleiades, in order to estimate their population ages and
metallicities. To this end we utilize a variety of photometric catalogues
covering the optical (Tycho $B_T$, $V_T$) and infrared (2MASS $J$, $K_s$),
in conjunction with a statistical analysis package, MATCH, which performs
the model-data comparison. The adopted distance moduli, extinction $A_{\rm{V}}$, 
and binary fractions used to model each cluster are collected in Table 
\ref{t:muav}. We fixed these values and assumed no error in them as they are 
fairly well determined for these clusters. Errors were calculated via 
analysis of the marginalized posterior probability distribution (as in \S
\ref{sssec:mock}) calculated through our fitting procedure (\S 
\ref{sssec:match}). In several cases, derived metallicities encounter the 
boundary of our search space ([Fe/H]$=0.40$). Hence, posterior probability 
distribution is not fully sampled in most cases. For these cases, we report 
the lower limit in Table \ref{t:results}, defined as the limit containing 
68\% of the probability in the posterior probability distribution.

\subsection{Fitting MESA models to the Hyades, Praesepe, and Pleiades with MATCH}
\label{ssec:fits}
We are exploring the effects that stellar rotation has on derived cluster 
age and metallicity. With fixed distance and extinction, we fit to the MS 
and MSTO of the Hyades, Praesepe, and Pleiades. This is performed through 
Hess diagrams comparing observed to synthetic magnitudes and colors. A 
pair of results is presented sequentially for each cluster below; derived 
assuming a single population rotation rate, or assuming a Gaussian 
distribution of rates (referenced as $P\left(\Omega/\Omega_{\rm{c}}=0.3\right)$; 
\S \ref{sssec:match}).

\begin{figure*}[!htb]
  \center
    \includegraphics[width=0.95\linewidth]{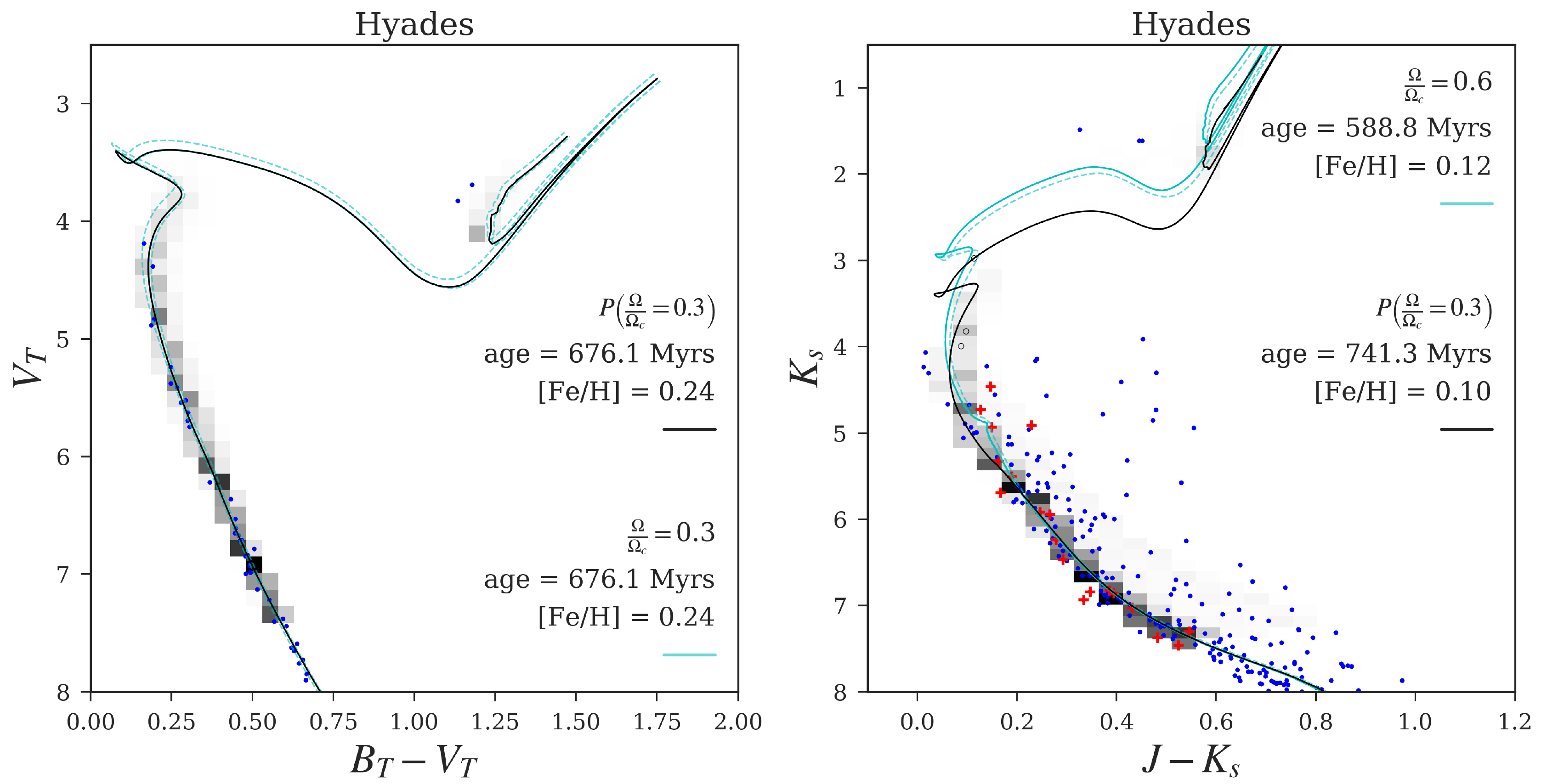}
    \caption{Best fit isochrones resulting from our analysis of $B_T$, $V_T$ 
      (from \citealt{deB2001}) is shown at left, and to 2MASS $J$, $K_s$ data 
      (from \citealt{BG2013}) of the Hyades at right. The best fitting 
      isochrone in absence of a rotation distribution is shown as a solid 
      turquoise line; dashed lines show isochrones at $\pm$ the uncertainty in 
      best fit age. The black solid line represents the best fit isochrone 
      from parameters derived via the best fit $P\left(\Omega/\Omega_{\rm{c}}=0.3\right)$ model (\S 
      \ref{sssec:match}); the rotation rate of this best fit isochrone is taken to be 0.3, i.e., 
      the $\Omega/\Omega_c$ value corresponding to the peak of the Gaussian distribution. Red crosses display the right hand side data 
      existent in the left hand side plot, matched on RA and DEC. Open 
      circles (only 3, near the upper MSTO in this figure) represent stars 
      excluded from the fit in MATCH in order to focus on the blueward MSTO 
      stars. The Hess diagram of the best fitting $P\left(\Omega/\Omega_{\rm{c}}=0.3\right)$
      model, is overplotted in gray to give a 
      sense of the MSTO spread from rotation effects}
\label{fig:hyadfits_rot}
\end{figure*}

\begin{figure*}[!htb]
  \center
    \includegraphics[width=0.95\linewidth]{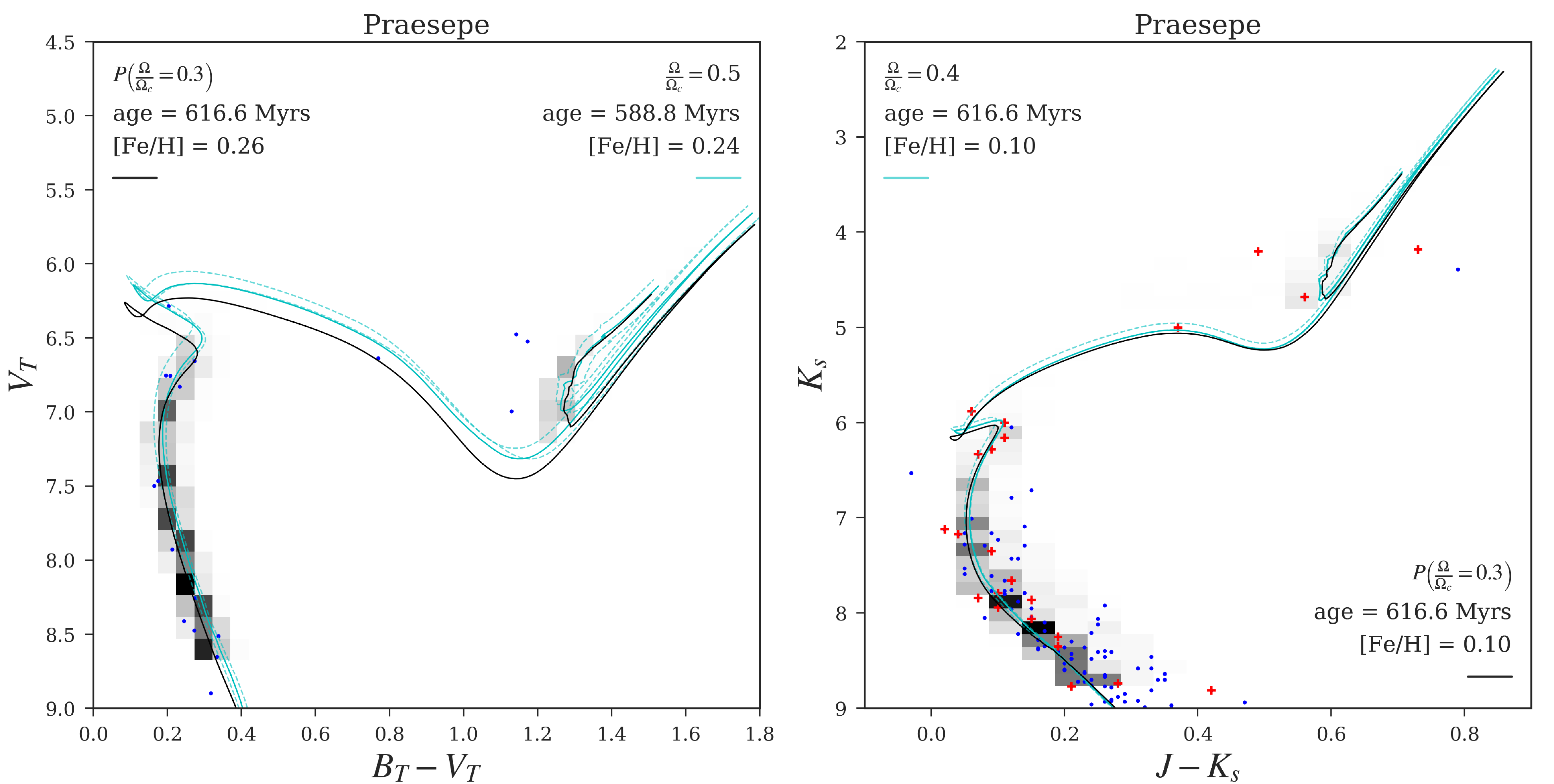}
    \caption{Similar to Figure \ref{fig:hyadfits_rot}, now for the Praesepe.
      The fit to $B_T$, $V_T$ (from members listed by \citealt{MDL2002}) is
      shown at left, and to 2MASS $J$, $K_s$ data (from \citealt{PFW2014}) 
      at right.}
\label{fig:praefits_rot}
\end{figure*}

\begin{figure*}[!htb]
  \center
    \includegraphics[width=0.95\linewidth]{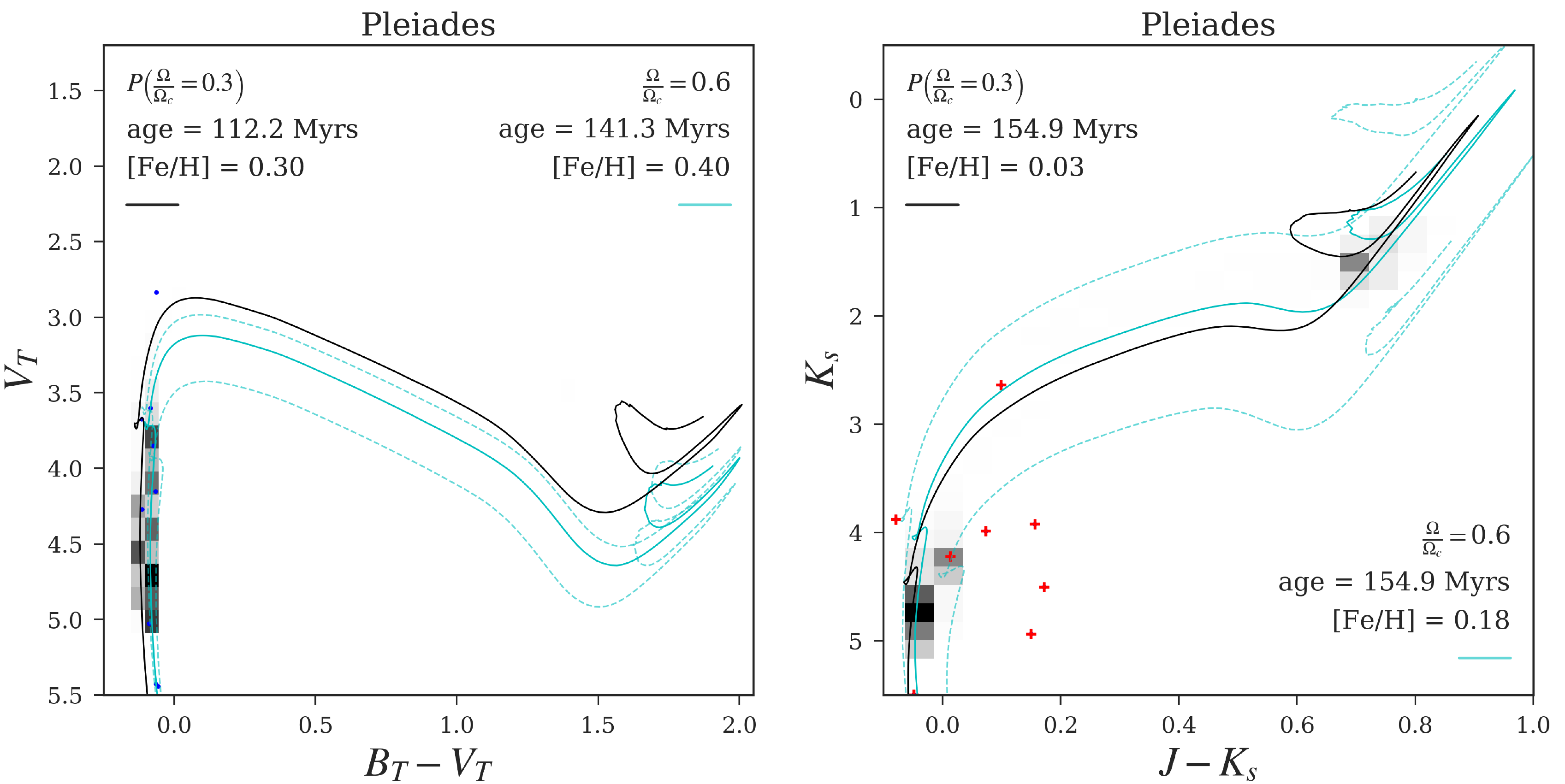}
    \caption{Similar to Figure \ref{fig:hyadfits_rot}, now for the
      Pleiades. The fit to $B_T$, $V_T$ (from members listed by
      \citealt{MDL2002}) is shown at left, and to 2MASS $J$, $K_s$
      data (from \citealt{JRS2007}) at right.}
\label{fig:pleifits_rot}
\end{figure*}

Figures \ref{fig:hyadfits_rot}, \ref{fig:praefits_rot}, and 
\ref{fig:pleifits_rot} show best fitting isochrones overlaid with observed 
data from the Hyades, Praesepe, and Pleiades. These are representative 
isochrones from our MESA models; representative in that they exist on model 
grid points, while the reported best fit values from MATCH belong to a 
continuum. The displayed isochrone ages are the closest grid point values to 
the MATCH reports. We have overlaid the Hess diagram of the best-fit 
$P\left(\Omega/\Omega_{\rm{c}}=0.3\right)$ model in these figures to give a sense of 
what our models truly look like (shown in gray). For instance, refer back to Figure \ref{fig:moddemo} for the general
appearance of our models, with the effects of gravity darkening and a 
distribution of rotation rates broadening the MSTO.

Best fit isochrones appear to trace the data qualitatively well. Although, 
in Figures \ref{fig:hyadfits_rot} and \ref{fig:praefits_rot}, corresponding 
to the Hyades and Praesepe, it may be seen that the red clump region in each
CMD is not fit very well, perhaps due to complexities in convection 
(see \S \ref{ssec:cov}). Teal isochrones correspond to fits assuming no 
distribution of rotation rates. In black, the fit utilizing 
$P\left(\Omega/\Omega_{\rm{c}}=0.3\right)$ is shown. Solid lines represent 
isochrones at the best fit values of age and metallicity. Dashed lines are 
the best fit isochrone at its $\pm$ uncertainty values in age; these are 
not displayed with $P\left(\Omega/\Omega_{\rm{c}}=0.3\right)$, for clarity. 
Derived errors and best fit values are collected in Table \ref{t:results}.

For the Hyades, we mostly find ages that are consistent with classical 
non-rotating CMD analyses, and the LDB age determined by \cite{ELM2018}, 
in both filter sets. Metallicities are within $\sim0.05$ dex of 
spectroscopic values. For instance, in $B_T$, $V_T$, the age is found to be 
$676^{+67}_{-11}$ Myr and [Fe/H]$=0.23^{+0.02}_{-0.02}$ when modeling a 
distribution of rotation rates. This age is nearer to results found using 
non-rotating models, although on the higher end. Ages of $\sim680$ Myr are 
found for the Hyades, whether using a distribution of rotation rates or a 
fixed value. This is true for the Hyades in all cases, except with our 
age determined via NIR data when modeling a distribution of rotation rates. 

For the Hyades in NIR, when using a Gaussian rotation distribution, we see 
an age of $741^{+55}_{-11}$ Myr. This older age may result from the greater 
color span of MSTO stars here. In some sense, this resembles a broadened 
MSTO; using a distribution of rotation rates also makes the MSTO more broad. 
Thus, the $P\left(\Omega/\Omega_{\rm{c}}=0.3\right)$ model can fit both the redder 
and bluer MSTO stars by sitting in the middle, driving an older age than the 
fixed $\Omega/\Omega_{\rm{c}}$ model, which is forced to fit either the redward or 
blueward MSTO stars moreso. On this point, note that our best fit single 
$\Omega/\Omega_{\rm{c}}$ model for the Hyades in NIR is high (0.6) compared to the 
lower value found in optical. This is due to the relatively low color span 
of MSTO stars in optical compared to NIR for the Hyades; the effects of 
gravity darkening produce a greater color spread in MSTO stars at higher 
rotation rates, driving the best fit model its higher $\Omega/\Omega_{\rm{c}}$ 
value. However, the MSTO spread in this case may be due to complications in 
the photometry. Membership and binarity uncertainties could be at play here, 
i.e., this older age could be spurious; (e.g., \citealt{TGK2016} find a 
tighter NIR MSTO morphology, identifying single stars in the Hyades). We 
excluded several of the redder MSTO stars from this fit, indicated in Figure 
\ref{fig:hyadfits_rot} by the open circles (there are 3 on the upper MSTO) 
in 2MASS $J$, $K_s$. This was done to reduce the influence of redward MSTO 
stars on derived age here. The Praesepe and Pleiades do not display as large 
of a color span in their MSTOs with NIR data.

It may also be noted that a much higher value of $\Omega/\Omega_c$ is 
preferred in fitting the NIR data of the Hyades (0.6 vs. 0.3 in optical). 
This is likely due to the MSTO of the Hyades having a greater spread in 
color and more MSTO stars in the NIR data. Adopting a higher 
$\Omega/\Omega_c$ grants models a broadened MSTO, via gravity darkening 
(e.g., Fig. \ref{fig:modgd}). We speculate that this allows a better fit to 
the NIR data, but is statistically worse in application to the Hyades 
optical morphology, which lacks a significant color spread in its MSTO. 
However, we leave a more thorough explanation of this discrepancy to future 
work.

Thus, for the Hyades, our ages mostly resemble those of classical 
non-rotating isochrone determinations, e.g., \cite{MAP1998}. The metallicity 
results roughly align with the measurement of [Fe/H] $\approx0.15$ from 
\cite{JDC2017}, although our results from the optical bands tend to favor 
values nearer to [Fe/H]$=0.20$, while in NIR we find nearer to [Fe/H]$=0.10$.

The situation for the Praesepe is similar to that of the Hyades: we see 
results nearer to reports from non-rotating models. This is the case across 
the NIR and optical data sets, whether using fixed $\Omega/\Omega_{\rm{c}}$ or 
$P\left(\Omega/\Omega_{\rm{c}}=0.3\right)$. Here, the ages tend towards $\sim590$ 
Myr, while the best fits for [Fe/H] are near $\sim0.09$ in the NIR and in 
$B_T$, $V_T$, we find [Fe/H]$\sim0.25$. Here our optical data 
set contains fewer MS stars than our NIR, making metallicity more difficult 
to determine, driving higher [Fe/H] values here.

For the Pleiades, we similarly find ages that agree with classical, 
non-rotating analyses, as well as LDB results. Our ages fall within the 
range $\sim112-160$ Myr, in concordance with values derived from the LDB, 
and non-rotating isochrones. In the NIR, using a distribution of 
rotation rates, the ages have a large uncertainties. The MSTO of the 
Pleiades in NIR is relatively more extended, driving this behavior. 
Values of [Fe/H] range from $-0.01$ to $0.40$ (our [Fe/H] search boundary); 
these tend to be higher than established values. Similar to the scenario 
seen with the optical data set for the Praesepe, the number of stars here is 
very low: only 7. Our magnitude cuts of $V_T < 5.0$ and $K_s < 5.0$ were 
made to focus more on the MSTO, at the expense of constraints from the 
numerous MS Pleiads. 


\begin{figure*}[!htb]
\center
\includegraphics[width=0.95\linewidth]{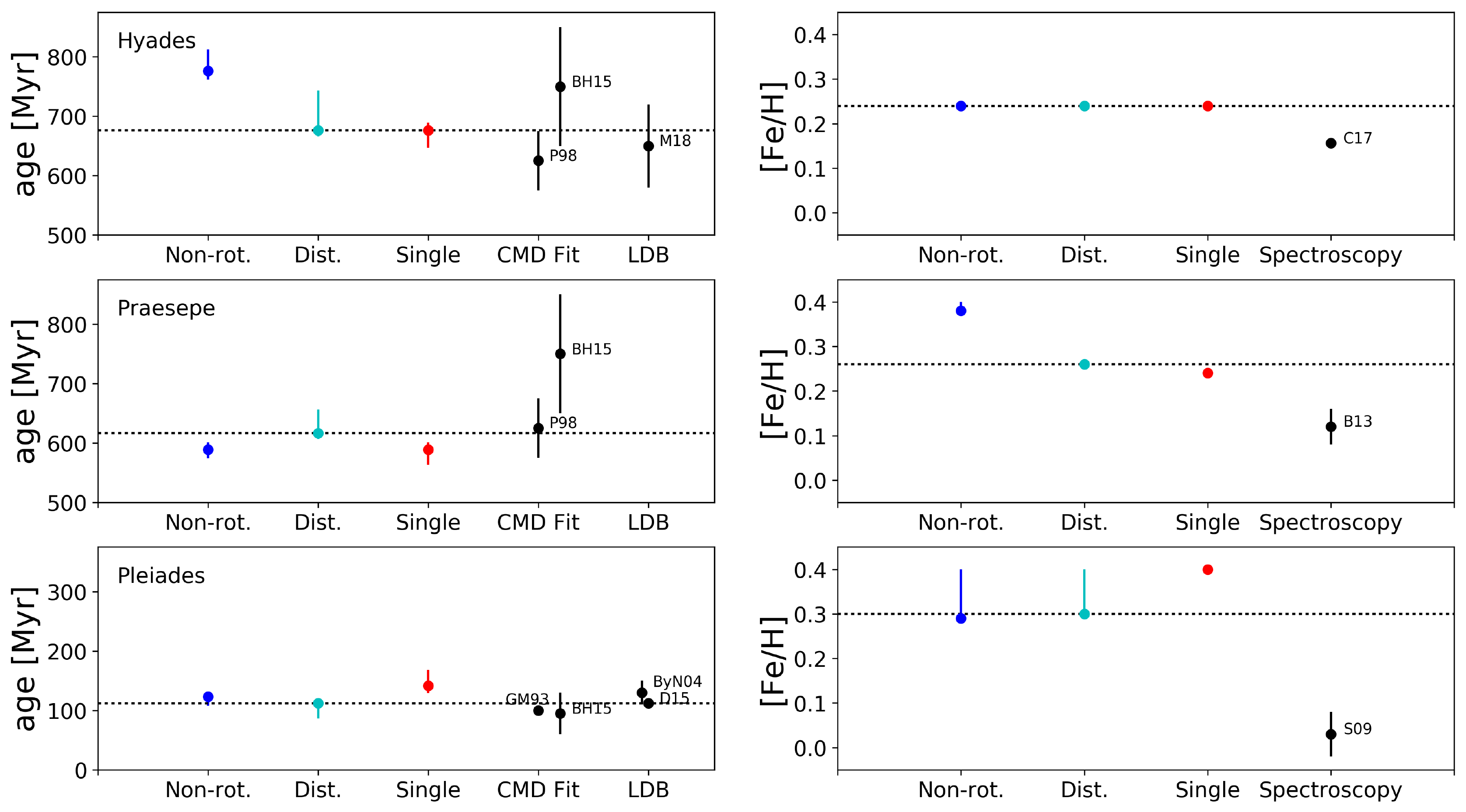}
\vspace{0.1cm}
\caption{Summary of results from Tycho $B_T$, $V_T$ photometry. Several 
  literature values are displayed in black, alongside our results 
  derived with and without a rotation rotation distribution turquoise and red, 
  resp. (see \S \ref{sssec:rotation} for details on the distribution) for 
  the population; non-rotating results are shown in blue. Literature 
  values above are \cite{MAP1998} (P98; used non-rotating isochrones), 
  \cite{GM1993} (GM93; used non-rotating isochrones), \cite{ELM2018} 
  (M18; used the LDB), \cite{ByN2004} (ByN04; used the LDB), 
  \cite{JDC2017} (C17; used spectra of 37 Hyads), \cite{AB2013} (B13; 
  used 11 Praesepe dwarfs), \cite{DS2009} (S09; from spectra of 20 
  Pleiads), \cite{SED2015} (D15; used the LDB). BH15 refers to 
  \cite{BH2015a} and \cite{BH2015b}, where rotating Geneva stellar models, 
  interpolated with non-rotating PARSEC models, were fit to MSTO stars of 
  these clusters. The black dotted line is included to aid comparisons of 
  our full model (gravity darkening and rotation distribution) to other 
  results. Unconstrained values in [Fe/H] are lower limits (see text), 
  given an uncertainty reaching to the boundary of [Fe/H] parameter 
  search space: 0.40.}
\label{fig:bvres}
\end{figure*}

\begin{figure*}[!htb]
\center
\includegraphics[width=0.95\linewidth]{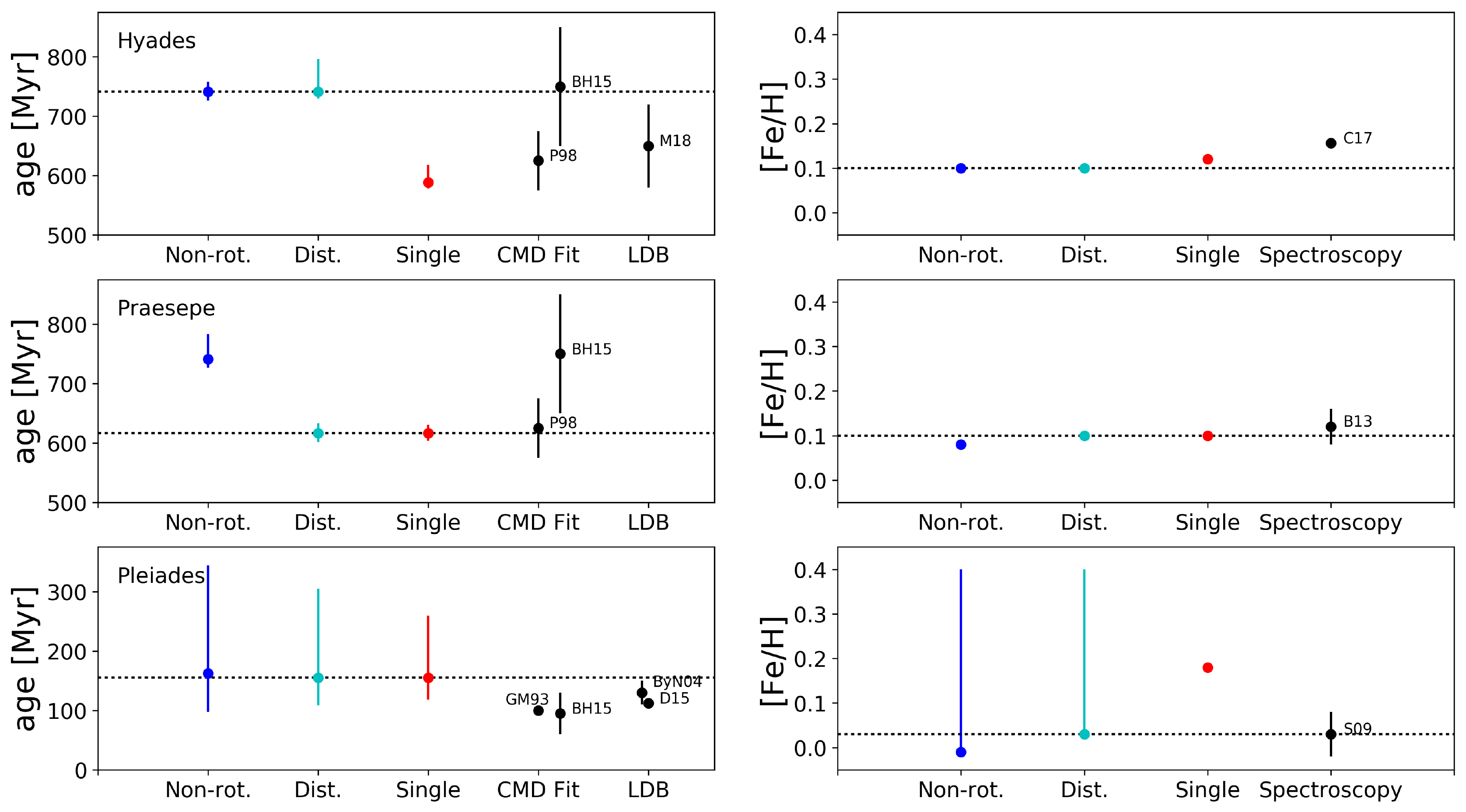}
\vspace{0.1cm}
\caption{Summary of results, but now for 2MASS photometry; see Figure 
         \ref{fig:bvres} for details. For a discussion on the higher 
         age derived here with the Hyades under a Gaussian distribution 
         of rotation rates, see \S \ref{ssec:fits}.}
\label{fig:2mres}
\end{figure*}


With the Pleiades, we did attempt fits with less severe magnitude cuts, but 
this in turn led to an increased age. On a Hess diagram, the low stellar 
density in the MSTO makes its stars appear as outliers in determining a 
global fit, despite these few stars being perhaps the most significant to 
extracting an accurate age. With relatively few stars, our uncertainties on 
age are larger for this cluster than the others. Moreover, many of the 
derived metallicities are unbounded and not well determined.

\clearpage

\section{Discussion}
\label{s:discussion}
Here we present a comparison of ages derived in previous literature to our
derived ages for the Hyades, Praesepe, and Pleiades in \S
\ref{ssec:complit}. Following this is a discussion on our results in
comparison to ages derived using rotating models built with the Geneva code,
and interpolated with PARSEC stellar models for a greater range in
metallicity and finer mass resolution (\S \ref{ssec:compgeneva}). Finally,
we discuss the effect of mixing length on the red clump region (\S
\ref{ssec:cov}). The red clump's CMD position is sensitive to the efficiency
of convection, and our default models were found to not match this region
well; altering the efficiency may be a possible solution, at least in some
cases.


\subsection{Comparison to Literature Ages and Metallicities}
\label{ssec:complit}
Figure \ref{fig:bvres} shows our best fit age and [Fe/H] from fitting to 
optical data in comparison to several literature values; Figure
\ref{fig:2mres} shows the same for our infrared data fits. For the Hyades,
Praesepe, and Pleiades, we generally find ages that align with classical 
results (although see \S \ref{ssec:fits} about the higher age derived for 
the Hyades in our NIR fit). Our metallicities for the Hyades, Praesepe, in 
the optical, and the Pleiades in both optical and infrared, appear to be 
inconsistent with literature values. Metallicity is unconstrained in these 
cases, leading to a dubious metallicity determination that lies near the end 
of our search space in metallicity: 0.40 dex. That metallicity appears 
unconstrained here is probably due to the relatively low stellar number 
densities present in these data sets, and the issues that this may cause 
with analysis on a Hess diagram. Additionally, we excluded much of the MS in 
our fits to the Pleiades, sacrificing its use in constraining metallicity, 
opting to focus on the sparse, brighter Pleiads for an age determination 
that is driven more by the MSTO.

The derived ages of the Hyades and Praesepe from CMD analysis using
non-rotating isochrones and the LDB congregate near $625-650$ Myr (see
references in \S \ref{ssec:hyades} and \S \ref{ssec:praesepe}). For the
Pleiades, non-rotating isochrone and LDB analysis have derived ages from
about $100-130$ Myr (see references in \S \ref{ssec:pleiades}). Our ages are 
consistent with classical non-rotating isochrone CMD fits and LDB results 
(where available) for the Hyades, Praesepe, and Pleiades.

Our results contrast with the ages derived by \cite{BH2015a} (see e.g., the 
points labeled BH15 in Figure \ref{fig:bvres} and \ref{fig:2mres}). Derived 
ages of the Hyades and Praesepe from CMD analysis using rotating Geneva 
isochrones, carried out by \cite{BH2015a, BH2015b}, using a Bayesian 
approach, are greater than classical reports. An age of $750\pm100$ Myr was 
derived for Hyades ($800\pm50$ if fixing metallicity to $[Fe/H]=0.10$). The 
age of the Praesepe was determined to be $790\pm60$ Myr. Lastly, an age of 
$\sim95\pm35$ Myr was determined for the Pleiades, consistent with classical 
non-rotating age determinations and the LDB. Our derived ages agree with 
these studies for the Pleiades, but we do not find as great of an age
increase for the Hyades nor the Praesepe via our models and methods.

The metallicity results for the Hyades and Praesepe are lower by $\sim0.05$ 
dex of literature values in our 2MASS fits; in $B_T$, $V_T$, best fit [Fe/H] 
for the Hyades and Praesepe are higher by $\sim0.05$ dex of literature 
values. Our metallicity results for the Pleiades are generally inconsistent 
with literature values (those being [Fe/H]$\approx0.0$); this is likely due 
to our magnitude cuts excluding much of the MS in favor of MSTO stars, and 
thus [Fe/H] becoming unconstrained in this CMD region. For instance, note 
the similarity in CMD position of the isochrones in Figure 
\ref{fig:pleifits_rot} (right panel), although they differ by 
$\sim0.10-0.15$ dex in metallicity.

Generally, rotating models are preferred owing to a mildly higher probability 
over non-rotating models. MATCH uses a Poisson equivalent to $\chi^2$ as a 
fit statistic (e.g., see \citealt{AD2001}, \S 2.3), designated via $-2\rm{ln}P$ 
(a lower value corresponds to higher probability), where $\rm{P}$ is the 
cumulative Poisson likelihood ratio, incorporating all Hess diagram bins. 
For the Hyades, we find $-2\rm{ln}P=104$ in optical photometry for the 
best fitting model, while the non-rotating model shows $-2\rm{ln}P=120$, 
for instance. The best-fit model with a Gaussian distribution of rotation 
rates achieved $-2\rm{ln}P=101$ in this case. Typically, rotating models 
are preferred according to the fit statistic by roughly $3-25$\% over 
non-rotating models. 

The results derived via models with a Gaussian distribution of 
$\Omega/\Omega_{\rm{c}}$ or a fixed value are often similar. However, as these 
clusters contain relatively low stellar densities in their MSTOs, they 
do not provide a strong distinction between more realistic models that 
include a distribution of rotation rates, and those that possess a fixed 
population $\Omega/\Omega_{\rm{c}}$ value. With precise photometry from upcoming 
data sets like $Gaia$ DR2, and in studying more populous clusters such as 
those in the LMC and SMC, we will have a better opportunity to assess 
the role of a rotation distribution.

\subsection{Comparison to Geneva Models}
\label{ssec:compgeneva}
Our models behave differently from Geneva models, as highlighted in \S 
\ref{sssec:rotation}, so it may be no surprise that our results differ from 
work that utilizes the Geneva models. Rotation has more modest consequences 
for stellar evolution under our physical assumptions, so we do not observe a 
strong affect on derived ages due to stellar rotation. No major differences 
in derived age appear to manifest from using a Gaussian distribution of 
rotation rates as opposed to a single value. Indeed, the MSTOs present in 
these clusters tend to not be especially well populated, and so any spread 
that may exist due to a distribution of rotation rates may be difficult to 
observe.

As rotation tends to make the MSTO primarily cooler in our modeling, at 
least at ages near $600$ Myr (see Figure \ref{fig:vvcvary}), it makes sense 
that we do not find a significantly older age for the Hyades or Praesepe. 
Rather, adopting a younger age would increase the brightness and make the 
MSTO hotter, helping an isochrone compensate for the effects incurred by an 
increased rotation rate. Vice versa in the case of the Pleiades at 
$\sim120$ Myr, here increasing rotation primarily makes the models brighter, 
leading models with higher $\Omega/\Omega_{\rm{c}}$ to mimic younger stars at fixed 
age\footnote{Note the slight increase in luminosity at the MSTO of the 
$M>2M_{\odot}$ models in Figure \ref{fig:trkcmp} and the more apparent 
enhancement in Figure \ref{fig:vvcvary} at $\sim120 Myr$.}.

This is in contrast to results found by \cite{BH2015a}, \cite{BH2015b}, 
where the Praesepe and Hyades showed an age increase of roughly 200 Myr as 
a result of stellar rotation in their modeling. In Geneva models, the
changes in isochrone morphology due to rotation are almost completely 
opposite to the effects seen in our models. Geneva models are hotter, 
brighter, and live longer on the MS (meaning the population evolves more 
slowly), as discussed in \S \ref{sssec:rotation}. It seems plausible that 
this discrepancy may stem from a more modest rotational mixing efficiency 
in MIST compared to Geneva. Whether this is precisely the case, or perhaps 
if other model differences come into play more strongly will require further 
investigation. It is also important to bear in mind that our model set is 
limited to $\Omega/\Omega_c=0.0$ to $0.6$ presently, while the Geneva models 
allow study from $\Omega/\Omega_c=0.0$ to $1$. Extending our model grid to 
include $\Omega/\Omega_c > 0.6$ will be fruitful for comparison, aside from 
perhaps being necessary to study realistic rotation distributions (e.g., those 
discussed in \citealt{HGMcS2010}); we aim to incorporate models with higher 
rotation in future work.

Furthermore, our models possess a greater mixing due to CCO. This was shown 
in Figure \ref{fig:trkcmp} to make our non-rotating models hotter and 
brighter than their Geneva counterparts. The effects of CCO mixing are 
similar to rotationally enhanced mixing, effectively expanding the stellar 
core, granting it more fuel to burn longer, brighter, and hotter. Our 
non-rotating age determinations are near what \cite{BH2015a} found with 
their rotating models (see Table \ref{t:results} and Figures \ref{fig:bvres} 
and \ref{fig:2mres}). It seems plausible that this is due to the higher 
level of CCO mixing in our models, as CCO makes the MSTO hotter and brighter. 
Proper 1D treatment of convection is yet another complication in stellar 
evolution theory and CMD-based analyses.

The physical assumptions made in our models \S \ref{ssec:models} and in 
Geneva are both able to simulate observational constraints. Much of our 
adopted physics follow from the assumptions made in MIST (\citealt{JC2016}; 
see \S 8 and 9 of that paper for comparisons to data), while the physics 
adopted in Geneva (\citealt{SE2012}; see \S 5 of that paper for comparisons 
to data) are similar, although there are differences, for example, in the treatment of 
rotational mixing and the assumed strength of convective core overshoot 
mixing. In our adopted formalism, the rotational mixing parameters $f_c$ 
and $f_{\mu}$ are tuned to match the observed nitrogen enrichment in 
galactic MS B-type stars (with the observations from e.g., 
\citealt{GL1992, K1992,MHB2008,IH2009}), following \cite{HLW2000}. The Geneva models are 
capable of reproducing these observed surface abundances without having 
any parameters calibrated to do so in their formalism, followed from 
\cite{MaZh1998}. Thus, although physical assumptions may differ between 
model sets, neither is ruled out by observational constraints thus far.

\subsection{Effect of Mixing Length on the Red Clump}
\label{ssec:cov}
\begin{figure}[!htb]
  \center
    \includegraphics[width=0.95\linewidth]{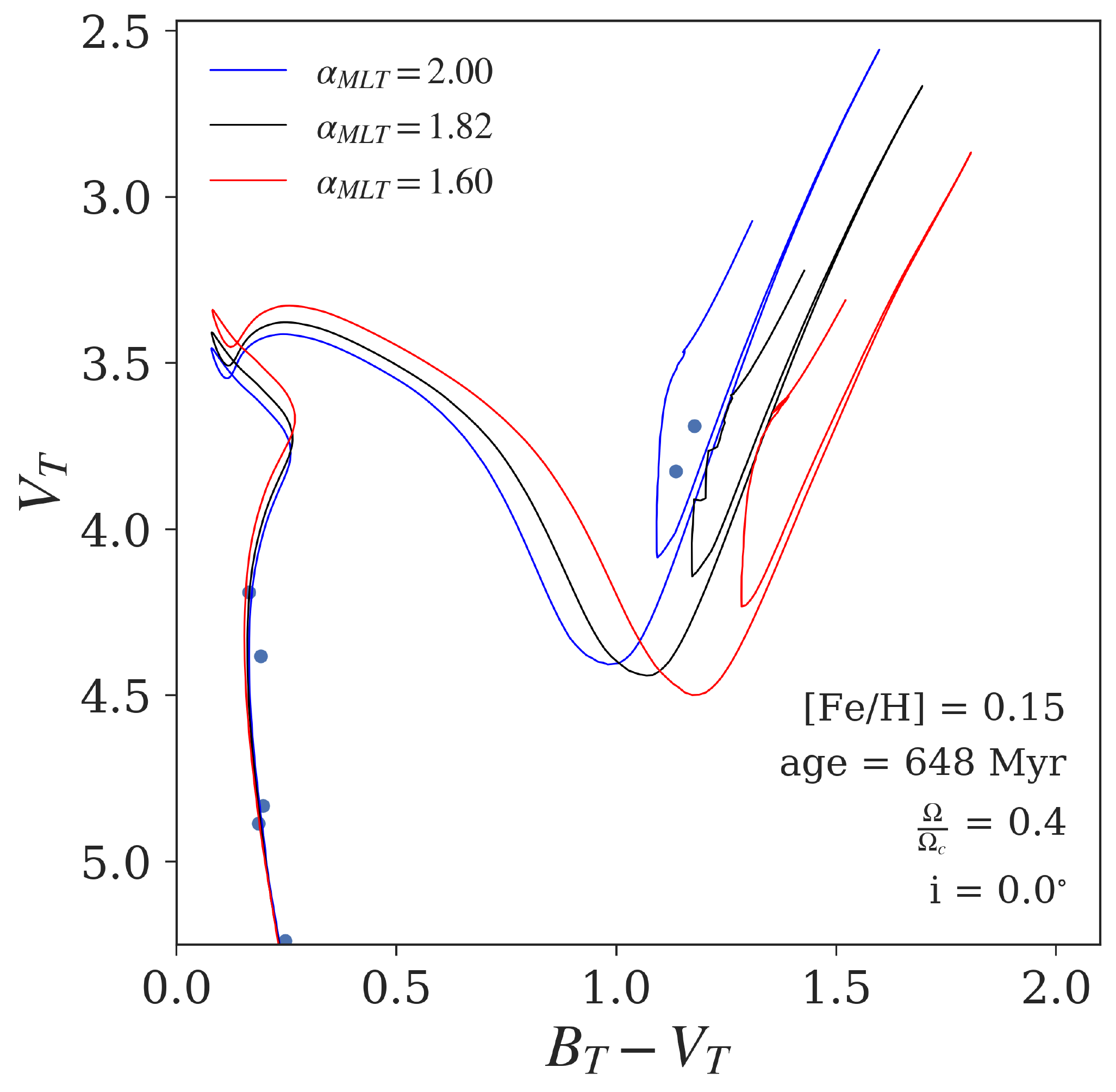}
    \caption{Showcasing the effect of altering the convective mixing length 
      parameter $\alpha_{\rm{MLT}}$. The default value of $\alpha_{\rm{MLT}}$ is $1.82$ 
      in our models (shown as black), calibrated to reproduce solar Li 
      surface abundances. Analogous isochrones are shown at 
      $\alpha_{\rm{MLT}}=1.60$ (red) and $\alpha_{\rm{MLT}}=2.00$ (blue) for comparison. 
      The data featured here is the cross matched data from the member list 
      of \citealt{deB2001} for the Hyades.}
\label{fig:amlt}
\end{figure}

Although our fits appear to match the observed MS and MSTO regions 
relatively well for these clusters, in Figures \ref{fig:hyadfits_rot} and 
\ref{fig:praefits_rot}, it is clear that our best-fit isochrones miss the 
red clump region. Figure \ref{fig:amlt} displays the range of convective 
mixing at various values, to demonstrate how it can affect the red clump 
region. In particular, we vary the parameter $\alpha_{\rm{MLT}}$, responsible for 
setting the enhanced range of convective mixing, according to the Mixing 
Length Theory (MLT) of \cite{BV1958}. This parameter essentially dictates 
how far a fluid parcel travels, $l_{\rm{MLT}}$, before mixing thermally with its 
surroundings; this length scale is characterized as some fraction of the 
pressure scale height, $H_{\rm{P}}$, and is related to the constant $\alpha_{\rm{MLT}}$ 
via $l_{\rm{MLT}}=\alpha_{\rm{MLT}}H_{\rm{P}}$.

By default $\alpha_{\rm{MLT}}=1.82$ in our models, based on solar calibration 
(discussed further in \citealt{JC2016}). Figure \ref{fig:amlt} shows isochrones, 
at age $649$ Myr (similar to our derived ages for the Hyades and Praesepe), 
[Fe/H]$=0.15$, $\Omega/\Omega_{\rm{c}}=0.4$, and inclination $i=0^{\circ}$, at 
various $\alpha_{\rm{MLT}}=1.60$, $1.82$ (default), and $2.00$. Altering 
$\alpha_{\rm{MLT}}$ makes the modeled red giant stars hotter or cooler, providing 
potential for a better fit to a given data set. In Figure \ref{fig:amlt} it 
is shown that a greater $\alpha_{\rm{MLT}}$ may provide a better fit in the
case of the Hyades. However, in the Praesepe this may lead to a worse fit, 
as it would pull the model away from two of the redder red clump stars.

Some 3D modeling efforts have seen evidence that $\alpha_{\rm{MLT}}$ may vary 
with e.g., stellar $T_{\rm{eff}}$, $\rm{log} g$, \citep{RT2015} and 
metallicity \citep{MWA2015}. Evidence that $\alpha_{\rm{MLT}}$ may vary with 
stellar parameters has also been seen in comparing models to observations, 
as in e.g., \cite{AB2012, JT2017, MJBC2018, S-HC2018, TL2018}, and 
\cite{LSV2018}. However, Choi et al. 2018 (submitted to ApJ) found that 
model $T_{\rm{eff}}$ can vary by nearly $\pm100$ K according to the treatment 
of surface boundary conditions in 1D codes. Thus, it appears that 
discrepancies between models and observations may (at least in part) be due 
to the chosen boundary conditions used in comparing models to data. 
Determining precisely how $\alpha_{\rm{MLT}}$ may vary, and according to which 
stellar properties, is an ongoing task. Efforts by e.g., \cite{WAEM2017} and 
\cite{JRM2017} will aid in having 3D simulations further inform our 1D 
models. Convection is another uncertain aspect in stellar modeling, 
alongside the uncertainties of stellar rotation. 

\section{Summary}
\label{s:summary}
In this paper our goal was to explore the effect of stellar rotation on the 
inferred ages of open clusters, using the well-studied examples of the 
Hyades, Praesepe, and Pleiades. Our results are summarized here:

\begin{itemize}
\item Application of self-consistent rotating isochrones (constructed via 
MESA) to CMD data of the Hyades, Praesepe, and Pleiades has yielded results 
that suggest no major influence on derived ages from the inclusion of 
rotation, even when including the effects of gravity darkening and a 
distribution of rotation rates (see \S \ref{ssec:models}). 

\item Our ages, derived using the statistical analysis package MATCH, are 
similar to values based on non-rotating models. Our models suggest ages of 
$\sim680$ Myr, $\sim590$ Myr, and $\sim110-160$ Myr for the Hyades, Praesepe, 
and Pleiades, respectively with our binary cleaned data. Our metallicity 
determinations roughly agree with literature in cases where the data 
provides a populous MS (i.e., in NIR for the Hyades and Praesepe, and the 
optical for the Hyades). These results are collected in Table 
\ref{t:results}.

\item These results are in contrast to the findings of \cite{BH2015a} who 
used Geneva stellar models, where a marked difference between rotating and 
non-rotating models was found for the Hyades and Praesepe clusters. In that 
work, rotation increased population ages by $\sim200$ Myr, as derived from 
the MSTO; our models find a less dramatic increase to derived age.
\end{itemize}

The physics that we have adopted in our modeling differs from what is 
adopted in Geneva. Either set of physics is well-founded in that they 
each are tuned to reproduce certain observations, and can do so 
successfully. Our models show comparably modest differences as rotation 
rate is varied; whereas the changes to stellar lifetime, luminosity, 
and temperature are more dramatic in the Geneva models (see 
\citealt{JC2016} for additional comparisons). Such model uncertainties 
complicate the establishment of an age for the Praesepe and Hyades based 
on CMD analysis. However, it is worth noting that our derived ages agree 
with what others have found via the LDB method, These results demonstrate 
a reason for caution in using rotating stellar models. We still contend 
with significant uncertainty in crucial evolutionary processes (particularly 
rotation and convection in this context), producing correspondingly 
uncertain results.

Moving forward, our models are primed to look at how stellar rotation may 
factor into the eMSTOs observed in LMC and SMC clusters. In future work, we 
plan to compile and present the predicted star formation histories of 
several such clusters using the models developed here. Gravity darkening and 
a distribution of rotation rates are able to significantly broaden the MSTO 
(e.g., \citealt{BdM2009} and our Figure \ref{fig:moddemo}). It may be 
that rotation is not able to fully explain for the observed eMSTO morphology 
(e.g., see \citealt{GGC2017}), but our upcoming work aims to assess the extent 
to which rotation can account for an MSTO spread, according to our models. 
Given alternate, yet still viable sets of physical assumptions exist in our 
models, we hope to further elucidate what stellar rotation may be capable of.

\section{Acknowledgments}
\label{s:ack}

We thank Timothy Brandt for helpful comments and discussion on earlier 
drafts of this paper. We would also like to thank the anonymous referee for 
their comments in improving the clarity of the paper. SG acknowledges the 
National Science Foundation Graduate Research Fellowship under grant No. 
DGE1745303. CC acknowledges support from NASA grant AST-1313280, and the 
Packard Foundation. This paper is based upon work supported by the National 
Aeronautics and Space Administration (NASA) under Contract No. NNG16PJ26C 
issued through the WFIRST Science Investigation Teams Program. Some of this 
material is based upon work supported by the National Science Foundation 
under Award No. 1501205. We would also like to thank Bill Paxton and the 
MESA community for making this work possible.


\begin{thebibliography}{}
\expandafter\ifx\csname natexlab\endcsname\relax\def\natexlab#1{#1}\fi

\bibitem[{{Abt} {et~al.}(1965){Abt}, {Barnes}, {Biggs}, \& {Osmer}}]{HAA1965}
{Abt}, H.~A., {Barnes}, R.~C., {Biggs}, E.~S., \& {Osmer}, P.~S. 1965, \apj,
  142, 1604

\bibitem[{{Arnett} \& {Moravveji}(2017)}]{WAEM2017}
{Arnett}, W.~D., \& {Moravveji}, E. 2017, \apjl, 836, L19

\bibitem[{{Asplund} {et~al.}(2009){Asplund}, {Grevesse}, {Sauval}, \&
  {Scott}}]{MA2009}
{Asplund}, M., {Grevesse}, N., {Sauval}, A.~J., \& {Scott}, P. 2009, \araa, 47,
  481

\bibitem[{{Aufdenberg} {et~al.}(2006){Aufdenberg}, {M{\'e}rand}, {Coud{\'e} du
  Foresto}, {Absil}, {Di Folco}, {Kervella}, {Ridgway}, {Berger}, {ten
  Brummelaar}, {McAlister}, {Sturmann}, {Sturmann}, \& {Turner}}]{JA2006}
{Aufdenberg}, J.~P., {M{\'e}rand}, A., {Coud{\'e} du Foresto}, V., {et~al.}
  2006, \apj, 645, 664

\bibitem[{{Baraffe} {et~al.}(1998){Baraffe}, {Chabrier}, {Allard}, \&
  {Hauschildt}}]{IB1998}
{Baraffe}, I., {Chabrier}, G., {Allard}, F., \& {Hauschildt}, P.~H. 1998, \aap,
  337, 403

\bibitem[{{Baraffe} {et~al.}(2015){Baraffe}, {Homeier}, {Allard}, \&
  {Chabrier}}]{IB2015}
{Baraffe}, I., {Homeier}, D., {Allard}, F., \& {Chabrier}, G. 2015, \aap, 577,
  A42

\bibitem[{{Barnes}(2007)}]{SB2007}
{Barnes}, S.~A. 2007, \apj, 669, 1167

\bibitem[{{Barrado y Navascu{\'e}s} {et~al.}(2004){Barrado y Navascu{\'e}s},
  {Stauffer}, \& {Jayawardhana}}]{ByN2004}
{Barrado y Navascu{\'e}s}, D., {Stauffer}, J.~R., \& {Jayawardhana}, R. 2004,
  \apj, 614, 386

\bibitem[{{Basri} {et~al.}(1996){Basri}, {Marcy}, \& {Graham}}]{BMG1996}
{Basri}, G., {Marcy}, G.~W., \& {Graham}, J.~R. 1996, \apj, 458, 600

\bibitem[{{Bastian} \& {de Mink}(2009)}]{BdM2009}
{Bastian}, N., \& {de Mink}, S.~E. 2009, \mnras, 398, L11

\bibitem[{{Bertelli} {et~al.}(2003){Bertelli}, {Nasi}, {Girardi}, {Chiosi},
  {Zoccali}, \& {Gallart}}]{GB2003}
{Bertelli}, G., {Nasi}, E., {Girardi}, L., {et~al.} 2003, \aj, 125, 770

\bibitem[{{Boesgaard} {et~al.}(2013){Boesgaard}, {Roper}, \& {Lum}}]{AB2013}
{Boesgaard}, A.~M., {Roper}, B.~W., \& {Lum}, M.~G. 2013, \apj, 775, 58

\bibitem[{{B{\"o}hm}(1963)}]{KHB1963}
{B{\"o}hm}, K.-H. 1963, \apj, 138, 297

\bibitem[{{B{\"o}hm-Vitense}(1958)}]{BV1958}
{B{\"o}hm-Vitense}, E. 1958, \zap, 46, 108

\bibitem[{{Bonaca} {et~al.}(2012){Bonaca}, {Tanner}, {Basu}, {Chaplin},
  {Metcalfe}, {Monteiro}, {Ballot}, {Bedding}, {Bonanno}, {Broomhall},
  {Bruntt}, {Campante}, {Christensen-Dalsgaard}, {Corsaro}, {Elsworth},
  {Garc{\'{\i}}a}, {Hekker}, {Karoff}, {Kjeldsen}, {Mathur}, {R{\'e}gulo},
  {Roxburgh}, {Stello}, {Trampedach}, {Barclay}, {Burke}, \&
  {Caldwell}}]{AB2012}
{Bonaca}, A., {Tanner}, J.~D., {Basu}, S., {et~al.} 2012, \apjl, 755, L12

\bibitem[{{Brandt} \& {Huang}(2015{\natexlab{a}})}]{BH2015a}
{Brandt}, T.~D., \& {Huang}, C.~X. 2015{\natexlab{a}}, \apj, 807, 58

\bibitem[{{Brandt} \& {Huang}(2015{\natexlab{b}})}]{BH2015c}
---. 2015{\natexlab{b}}, \apj, 807, 25

\bibitem[{{Brandt} \& {Huang}(2015{\natexlab{c}})}]{BH2015b}
---. 2015{\natexlab{c}}, \apj, 807, 24

\bibitem[{{Brott} {et~al.}(2011){Brott}, {de Mink}, {Cantiello}, {Langer}, {de
  Koter}, {Evans}, {Hunter}, {Trundle}, \& {Vink}}]{IB2011}
{Brott}, I., {de Mink}, S.~E., {Cantiello}, M., {et~al.} 2011, \aap, 530, A115

\bibitem[{{Chaboyer} \& {Zahn}(1992)}]{CZ1992}
{Chaboyer}, B., \& {Zahn}, J.-P. 1992, \aap, 253, 173

\bibitem[{{Choi} {et~al.}(2017){Choi}, {Conroy}, \& {Byler}}]{CCB2017}
{Choi}, J., {Conroy}, C., \& {Byler}, N. 2017, \apj, 838, 159

\bibitem[{{Choi} {et~al.}(2016){Choi}, {Dotter}, {Conroy}, {Cantiello},
  {Paxton}, \& {Johnson}}]{JC2016}
{Choi}, J., {Dotter}, A., {Conroy}, C., {et~al.} 2016, \apj, 823, 102

\bibitem[{{Chun} {et~al.}(2018){Chun}, {Yoon}, {Jung}, {Kim}, \&
  {Kim}}]{S-HC2018}
{Chun}, S.-H., {Yoon}, S.-C., {Jung}, M.-K., {Kim}, D.~U., \& {Kim}, J. 2018,
  \apj, 853, 79

\bibitem[{{Conroy}(2013)}]{CC2013}
{Conroy}, C. 2013, \araa, 51, 393

\bibitem[{{Correnti} {et~al.}(2015){Correnti}, {Goudfrooij}, {Puzia}, \& {de
  Mink}}]{MC2015}
{Correnti}, M., {Goudfrooij}, P., {Puzia}, T.~H., \& {de Mink}, S.~E. 2015,
  \mnras, 450, 3054

\bibitem[{{Cummings} {et~al.}(2017){Cummings}, {Deliyannis}, {Maderak}, \&
  {Steinhauer}}]{JDC2017}
{Cummings}, J.~D., {Deliyannis}, C.~P., {Maderak}, R.~M., \& {Steinhauer}, A.
  2017, \aj, 153, 128

\bibitem[{{Dahm}(2015)}]{SED2015}
{Dahm}, S.~E. 2015, \apj, 813, 108

\bibitem[{{de Bruijne} {et~al.}(2001){de Bruijne}, {Hoogerwerf}, \& {de
  Zeeuw}}]{deB2001}
{de Bruijne}, J.~H.~J., {Hoogerwerf}, R., \& {de Zeeuw}, P.~T. 2001, \aap, 367,
  111

\bibitem[{{de Jong} {et~al.}(2008){de Jong}, {Rix}, {Martin}, {Zucker},
  {Dolphin}, {Bell}, {Belokurov}, \& {Evans}}]{JdJ2008}
{de Jong}, J.~T.~A., {Rix}, H.-W., {Martin}, N.~F., {et~al.} 2008, \aj, 135,
  1361

\bibitem[{{Dolphin}(1997)}]{AD1997}
{Dolphin}, A. 1997, New Astronomy, 2, 397

\bibitem[{{Dolphin}(2002)}]{AD2001}
{Dolphin}, A.~E. 2002, \mnras, 332, 91

\bibitem[{{Dotter}(2016)}]{ADo2016}
{Dotter}, A. 2016, \apjs, 222, 8

\bibitem[{{Ekstr{\"o}m} {et~al.}(2012){Ekstr{\"o}m}, {Georgy}, {Eggenberger},
  {Meynet}, {Mowlavi}, {Wyttenbach}, {Granada}, {Decressin}, {Hirschi},
  {Frischknecht}, {Charbonnel}, \& {Maeder}}]{SE2012}
{Ekstr{\"o}m}, S., {Georgy}, C., {Eggenberger}, P., {et~al.} 2012, \aap, 537,
  A146

\bibitem[{{Endal} \& {Sofia}(1978)}]{ES1978}
{Endal}, A.~S., \& {Sofia}, S. 1978, \apj, 220, 279

\bibitem[{{Espinosa Lara} \& {Rieutord}(2011)}]{ELR2011}
{Espinosa Lara}, F., \& {Rieutord}, M. 2011, \aap, 533, A43

\bibitem[{{Gaia Collaboration} {et~al.}(2016){Gaia Collaboration}, {Brown},
  {Vallenari}, {Prusti}, {de Bruijne}, {Mignard}, {Drimmel}, {Babusiaux},
  {Bailer-Jones}, {Bastian}, \& et~al.}]{GAIA2016}
{Gaia Collaboration}, {Brown}, A.~G.~A., {Vallenari}, A., {et~al.} 2016, \aap,
  595, A2

\bibitem[{{Gaia Collaboration} {et~al.}(2017){Gaia Collaboration}, {van
  Leeuwen}, {Vallenari}, {Jordi}, {Lindegren}, {Bastian}, {Prusti}, {de
  Bruijne}, {Brown}, {Babusiaux}, \& et~al.}]{GAIA2017}
{Gaia Collaboration}, {van Leeuwen}, F., {Vallenari}, A., {et~al.} 2017, \aap,
  601, A19

\bibitem[{{Gallart} {et~al.}(2003){Gallart}, {Zoccali}, {Bertelli}, {Chiosi},
  {Demarque}, {Girardi}, {Nasi}, {Woo}, \& {Yi}}]{CG2003}
{Gallart}, C., {Zoccali}, M., {Bertelli}, G., {et~al.} 2003, \aj, 125, 742

\bibitem[{{G{\'a}sp{\'a}r} {et~al.}(2009){G{\'a}sp{\'a}r}, {Rieke}, {Su},
  {Balog}, {Trilling}, {Muzzerole}, {Apai}, \& {Kelly}}]{AG2009}
{G{\'a}sp{\'a}r}, A., {Rieke}, G.~H., {Su}, K.~Y.~L., {et~al.} 2009, \apj, 697,
  1578

\bibitem[{{Gies} \& {Lambert}(1992)}]{GL1992}
{Gies}, D.~R., \& {Lambert}, D.~L. 1992, \apj, 387, 673

\bibitem[{{Girardi} {et~al.}(2002){Girardi}, {Bertelli}, {Bressan}, {Chiosi},
  {Groenewegen}, {Marigo}, {Salasnich}, \& {Weiss}}]{LG2002}
{Girardi}, L., {Bertelli}, G., {Bressan}, A., {et~al.} 2002, \aap, 391, 195

\bibitem[{{Girardi} {et~al.}(2000){Girardi}, {Bressan}, {Bertelli}, \&
  {Chiosi}}]{LG2000}
{Girardi}, L., {Bressan}, A., {Bertelli}, G., \& {Chiosi}, C. 2000, \aaps, 141,
  371

\bibitem[{{Girardi} {et~al.}(2013){Girardi}, {Goudfrooij}, {Kalirai}, {Kerber},
  {Kozhurina-Platais}, {Rubele}, {Bressan}, {Chandar}, {Marigo}, {Platais}, \&
  {Puzia}}]{LG2013}
{Girardi}, L., {Goudfrooij}, P., {Kalirai}, J.~S., {et~al.} 2013, \mnras, 431,
  3501

\bibitem[{{Goldman} {et~al.}(2013){Goldman}, {R{\"o}ser}, {Schilbach},
  {Magnier}, {Olczak}, {Henning}, {Juri{\'c}}, {Schlafly}, {Chen}, {Platais},
  {Burgett}, {Hodapp}, {Heasley}, {Kudritzki}, {Morgan}, {Price}, {Tonry}, \&
  {Wainscoat}}]{BG2013}
{Goldman}, B., {R{\"o}ser}, S., {Schilbach}, E., {et~al.} 2013, \aap, 559, A43

\bibitem[{{Goudfrooij} {et~al.}(2017){Goudfrooij}, {Girardi}, \&
  {Correnti}}]{GGC2017}
{Goudfrooij}, P., {Girardi}, L., \& {Correnti}, M. 2017, \apj, 846, 22

\bibitem[{{Goudfrooij} {et~al.}(2014){Goudfrooij}, {Girardi},
  {Kozhurina-Platais}, {Kalirai}, {Platais}, {Puzia}, {Correnti}, {Bressan},
  {Chandar}, {Kerber}, {Marigo}, \& {Rubele}}]{PG2014}
{Goudfrooij}, P., {Girardi}, L., {Kozhurina-Platais}, V., {et~al.} 2014, \apj,
  797, 35

\bibitem[{{Gouliermis} {et~al.}(2011){Gouliermis}, {Dolphin}, {Robberto},
  {Gruendl}, {Chu}, {Gennaro}, {Henning}, {Rosa}, {Da Rio}, {Brandner},
  {Romaniello}, {De Marchi}, {Panagia}, \& {Zinnecker}}]{DG2011}
{Gouliermis}, D.~A., {Dolphin}, A.~E., {Robberto}, M., {et~al.} 2011, \apj,
  738, 137

\bibitem[{{Guenther} {et~al.}(1992){Guenther}, {Demarque}, {Kim}, \&
  {Pinsonneault}}]{DBG1992}
{Guenther}, D.~B., {Demarque}, P., {Kim}, Y.-C., \& {Pinsonneault}, M.~H. 1992,
  \apj, 387, 372

\bibitem[{{Gunn} {et~al.}(1988){Gunn}, {Griffin}, {Griffin}, \&
  {Zimmerman}}]{JEG1988}
{Gunn}, J.~E., {Griffin}, R.~F., {Griffin}, R.~E.~M., \& {Zimmerman}, B.~A.
  1988, \aj, 96, 198

\bibitem[{{Heger} \& {Langer}(2000)}]{HL2000}
{Heger}, A., \& {Langer}, N. 2000, \apj, 544, 1016

\bibitem[{{Heger} {et~al.}(2000){Heger}, {Langer}, \& {Woosley}}]{HLW2000}
{Heger}, A., {Langer}, N., \& {Woosley}, S.~E. 2000, \apj, 528, 368

\bibitem[{{Herrero}(1993)}]{AH1993}
{Herrero}, A. 1993, \ssr, 66, 137

\bibitem[{{Huang} {et~al.}(2010){Huang}, {Gies}, \& {McSwain}}]{HGMcS2010}
{Huang}, W., {Gies}, D.~R., \& {McSwain}, M.~V. 2010, \apj, 722, 605

\bibitem[{{Hunter} {et~al.}(2009){Hunter}, {Brott}, {Langer}, {Lennon},
  {Dufton}, {Howarth}, {Ryans}, {Trundle}, {Evans}, {de Koter}, \&
  {Smartt}}]{IH2009}
{Hunter}, I., {Brott}, I., {Langer}, N., {et~al.} 2009, \aap, 496, 841

\bibitem[{{Jeffery} {et~al.}(2016){Jeffery}, {von Hippel}, {van Dyk},
  {Stenning}, {Robinson}, {Stein}, \& {Jefferys}}]{EJJ2016}
{Jeffery}, E.~J., {von Hippel}, T., {van Dyk}, D.~A., {et~al.} 2016, \apj, 828,
  79

\bibitem[{{J{\o}rgensen} \& {Lindegren}(2005)}]{JL2005}
{J{\o}rgensen}, B.~R., \& {Lindegren}, L. 2005, \aap, 436, 127

\bibitem[{{Joyce} \& {Chaboyer}(2018)}]{MJBC2018}
{Joyce}, M., \& {Chaboyer}, B. 2018, \apj, 856, 10

\bibitem[{{Kalirai} \& {Tosi}(2004)}]{JKMT2004}
{Kalirai}, J.~S., \& {Tosi}, M. 2004, \mnras, 351, 649

\bibitem[{{Katahira} {et~al.}(1996){Katahira}, {Hirata}, {Ito}, {Katoh},
  {Ballereau}, \& {Chauville}}]{JIK1996}
{Katahira}, J.-I., {Hirata}, R., {Ito}, M., {et~al.} 1996, \pasj, 48, 317

\bibitem[{{Kilian}(1992)}]{K1992}
{Kilian}, J. 1992, \aap, 262, 171

\bibitem[{{Kippenhahn} \& {Thomas}(1970)}]{KT1970}
{Kippenhahn}, R., \& {Thomas}, H.-C. 1970, in IAU Colloq. 4: Stellar Rotation,
  ed. A.~{Slettebak}, 20

\bibitem[{{Kopytova} {et~al.}(2016){Kopytova}, {Brandner}, {Tognelli}, {Prada
  Moroni}, {Da Rio}, {R{\"o}ser}, \& {Schilbach}}]{TGK2016}
{Kopytova}, T.~G., {Brandner}, W., {Tognelli}, E., {et~al.} 2016, \aap, 585, A7

\bibitem[{{Kroupa}(2001)}]{PK2001}
{Kroupa}, P. 2001, \mnras, 322, 231

\bibitem[{{Kurucz}(1970)}]{RLK1970}
{Kurucz}, R.~L. 1970, SAO Special Report, 309

\bibitem[{{Kurucz}(1993)}]{RLK1993}
---. 1993, {SYNTHE spectrum synthesis programs and line data}

\bibitem[{{Levesque} {et~al.}(2012){Levesque}, {Leitherer}, {Ekstrom},
  {Meynet}, \& {Schaerer}}]{EML2012}
{Levesque}, E.~M., {Leitherer}, C., {Ekstrom}, S., {Meynet}, G., \& {Schaerer},
  D. 2012, \apj, 751, 67

\bibitem[{{Li} {et~al.}(2018){Li}, {Bedding}, {Huber}, {Ball}, {Stello},
  {Murphy}, \& {Bland-Hawthorn}}]{TL2018}
{Li}, T., {Bedding}, T.~R., {Huber}, D., {et~al.} 2018, \mnras, 475, 981

\bibitem[{{Lindegren} {et~al.}(2000){Lindegren}, {Madsen}, \&
  {Dravins}}]{LL2000}
{Lindegren}, L., {Madsen}, S., \& {Dravins}, D. 2000, \aap, 356, 1119

\bibitem[{{Lucy}(1967)}]{LBL1967}
{Lucy}, L.~B. 1967, \zap, 65, 89

\bibitem[{{Madsen} {et~al.}(2002){Madsen}, {Dravins}, \& {Lindegren}}]{MDL2002}
{Madsen}, S., {Dravins}, D., \& {Lindegren}, L. 2002, \aap, 381, 446

\bibitem[{{Maeder}(1975)}]{AM1975}
{Maeder}, A. 1975, \aap, 43, 61

\bibitem[{{Maeder}(1997)}]{AM1997}
---. 1997, \aap, 321, 134

\bibitem[{{Maeder}(1999)}]{AM1999}
---. 1999, \aap, 347, 185

\bibitem[{{Maeder} \& {Meynet}(2000{\natexlab{a}})}]{MaMe2000a}
{Maeder}, A., \& {Meynet}, G. 2000{\natexlab{a}}, \aap, 361, 159

\bibitem[{{Maeder} \& {Meynet}(2000{\natexlab{b}})}]{MaMe2000b}
---. 2000{\natexlab{b}}, \araa, 38, 143

\bibitem[{{Maeder} \& {Zahn}(1998)}]{MaZh1998}
{Maeder}, A., \& {Zahn}, J.-P. 1998, \aap, 334, 1000

\bibitem[{{Magic} {et~al.}(2010){Magic}, {Serenelli}, {Weiss}, \&
  {Chaboyer}}]{ZM2010}
{Magic}, Z., {Serenelli}, A., {Weiss}, A., \& {Chaboyer}, B. 2010, \apj, 718,
  1378

\bibitem[{{Magic} {et~al.}(2015){Magic}, {Weiss}, \& {Asplund}}]{MWA2015}
{Magic}, Z., {Weiss}, A., \& {Asplund}, M. 2015, \aap, 573, A89

\bibitem[{{Mamajek} \& {Hillenbrand}(2008)}]{MH2008}
{Mamajek}, E.~E., \& {Hillenbrand}, L.~A. 2008, \apj, 687, 1264

\bibitem[{{Mart{\'{\i}}n} {et~al.}(2018){Mart{\'{\i}}n}, {Lodieu}, {Pavlenko},
  \& {B{\'e}jar}}]{ELM2018}
{Mart{\'{\i}}n}, E.~L., {Lodieu}, N., {Pavlenko}, Y., \& {B{\'e}jar}, V.~J.~S.
  2018, ArXiv e-prints, arXiv:1802.07155

\bibitem[{{Mazzei} \& {Pigatto}(1988)}]{MP1988}
{Mazzei}, P., \& {Pigatto}, L. 1988, \aap, 193, 148

\bibitem[{{McAlister} {et~al.}(2005){McAlister}, {ten Brummelaar}, {Gies},
  {Huang}, {Bagnuolo}, {Shure}, {Sturmann}, {Sturmann}, {Turner}, {Taylor},
  {Berger}, {Baines}, {Grundstrom}, {Ogden}, {Ridgway}, \& {van
  Belle}}]{HAM2005}
{McAlister}, H.~A., {ten Brummelaar}, T.~A., {Gies}, D.~R., {et~al.} 2005,
  \apj, 628, 439

\bibitem[{{Mermilliod}(1981)}]{JM1981}
{Mermilliod}, J.~C. 1981, \aap, 97, 235

\bibitem[{{Meynet} \& {Maeder}(1997)}]{MeMa1997}
{Meynet}, G., \& {Maeder}, A. 1997, \aap, 321, 465

\bibitem[{{Meynet} \& {Maeder}(2000)}]{MeMa2000}
---. 2000, \aap, 361, 101

\bibitem[{{Meynet} {et~al.}(1993){Meynet}, {Mermilliod}, \& {Maeder}}]{GM1993}
{Meynet}, G., {Mermilliod}, J.-C., \& {Maeder}, A. 1993, \aaps, 98, 477

\bibitem[{{Morel} {et~al.}(2008){Morel}, {Hubrig}, \& {Briquet}}]{MHB2008}
{Morel}, T., {Hubrig}, S., \& {Briquet}, M. 2008, \aap, 481, 453

\bibitem[{{Mosumgaard} {et~al.}(2017){Mosumgaard}, {Silva Aguirre}, {Weiss},
  {Christensen-Dalsgaard}, \& {Trampedach}}]{JRM2017}
{Mosumgaard}, J.~R., {Silva Aguirre}, V., {Weiss}, A., {Christensen-Dalsgaard},
  J., \& {Trampedach}, R. 2017, in European Physical Journal Web of
  Conferences, Vol. 160, European Physical Journal Web of Conferences, 03009

\bibitem[{{Nelson} {et~al.}(1993){Nelson}, {Rappaport}, \& {Chiang}}]{NRC1993}
{Nelson}, L.~A., {Rappaport}, S., \& {Chiang}, E. 1993, \apj, 413, 364

\bibitem[{{Nemravov{\'a}} {et~al.}(2010){Nemravov{\'a}}, {Harmanec},
  {Kub{\'a}t}, {Koubsk{\'y}}, {Iliev}, {Yang}, {Ribeiro}, {{\v S}lechta},
  {Kotkov{\'a}}, {Wolf}, \& {{\v S}koda}}]{JN2010}
{Nemravov{\'a}}, J., {Harmanec}, P., {Kub{\'a}t}, J., {et~al.} 2010, \aap, 516,
  A80

\bibitem[{{Paxton} {et~al.}(2011){Paxton}, {Bildsten}, {Dotter}, {Herwig},
  {Lesaffre}, \& {Timmes}}]{BP2011}
{Paxton}, B., {Bildsten}, L., {Dotter}, A., {et~al.} 2011, \apjs, 192, 3

\bibitem[{{Paxton} {et~al.}(2013){Paxton}, {Cantiello}, {Arras}, {Bildsten},
  {Brown}, {Dotter}, {Mankovich}, {Montgomery}, {Stello}, {Timmes}, \&
  {Townsend}}]{BP2013}
{Paxton}, B., {Cantiello}, M., {Arras}, P., {et~al.} 2013, \apjs, 208, 4

\bibitem[{{Paxton} {et~al.}(2015){Paxton}, {Marchant}, {Schwab}, {Bauer},
  {Bildsten}, {Cantiello}, {Dessart}, {Farmer}, {Hu}, {Langer}, {Townsend},
  {Townsley}, \& {Timmes}}]{BP2015}
{Paxton}, B., {Marchant}, P., {Schwab}, J., {et~al.} 2015, \apjs, 220, 15

\bibitem[{{Paxton} {et~al.}(2018){Paxton}, {Schwab}, {Bauer}, {Bildsten},
  {Blinnikov}, {Duffell}, {Farmer}, {Goldberg}, {Marchant}, {Sorokina},
  {Thoul}, {Townsend}, \& {Timmes}}]{BP2018}
{Paxton}, B., {Schwab}, J., {Bauer}, E.~B., {et~al.} 2018, \apjs, 234, 34

\bibitem[{{Pearce} \& {Hill}(1971)}]{JP1971}
{Pearce}, J.~A., \& {Hill}, G. 1971, \pasp, 83, 493

\bibitem[{{Penny}(1996)}]{P1996}
{Penny}, L.~R. 1996, \apj, 463, 737

\bibitem[{{Perrin} {et~al.}(1977){Perrin}, {Cayrel de Strobel}, {Cayrel}, \&
  {Hejlesen}}]{PCdSC1977}
{Perrin}, M.-N., {Cayrel de Strobel}, G., {Cayrel}, R., \& {Hejlesen}, P.~M.
  1977, \aap, 54, 779

\bibitem[{{Perryman} {et~al.}(1997){Perryman}, {Brown}, {Lebreton}, {Gomez},
  {Turon}, {Cayrel de Strobel}, {Mermilliod}, {Robichon}, {Kovalevsky}, \&
  {Crifo}}]{MAP1997}
{Perryman}, M.~A.~C., {Brown}, A.~G.~A., {Lebreton}, Y., {et~al.} 1997, VizieR
  Online Data Catalog, 333

\bibitem[{{Perryman} {et~al.}(1998){Perryman}, {Brown}, {Lebreton}, {Gomez},
  {Turon}, {Cayrel de Strobel}, {Mermilliod}, {Robichon}, {Kovalevsky}, \&
  {Crifo}}]{MAP1998}
---. 1998, \aap, 331, 81

\bibitem[{{Piatti} \& {Cole}(2017)}]{PC2017}
{Piatti}, A.~E., \& {Cole}, A. 2017, ArXiv e-prints, arXiv:1705.08186

\bibitem[{{Pinsonneault} {et~al.}(1989){Pinsonneault}, {Kawaler}, {Sofia}, \&
  {Demarque}}]{MHP1989}
{Pinsonneault}, M.~H., {Kawaler}, S.~D., {Sofia}, S., \& {Demarque}, P. 1989,
  \apj, 338, 424

\bibitem[{{Ram{\'{\i}}rez-Agudelo} {et~al.}(2013){Ram{\'{\i}}rez-Agudelo},
  {Sim{\'o}n-D{\'{\i}}az}, {Sana}, {de Koter}, {Sab{\'{\i}}n-Sanjul{\'{\i}}an},
  {de Mink}, {Dufton}, {Gr{\"a}fener}, {Evans}, {Herrero}, {Langer}, {Lennon},
  {Ma{\'{\i}}z Apell{\'a}niz}, {Markova}, {Najarro}, {Puls}, {Taylor}, \&
  {Vink}}]{RA2013}
{Ram{\'{\i}}rez-Agudelo}, O.~H., {Sim{\'o}n-D{\'{\i}}az}, S., {Sana}, H.,
  {et~al.} 2013, \aap, 560, A29

\bibitem[{{Reino} {et~al.}(2018){Reino}, {de Bruijne}, {Zari}, {d'Antona}, \&
  {Ventura}}]{SR2018}
{Reino}, S., {de Bruijne}, J., {Zari}, E., {d'Antona}, F., \& {Ventura}, P.
  2018, \mnras, arXiv:1804.00759

\bibitem[{{Richichi} {et~al.}(1994){Richichi}, {Calamai}, \&
  {Leinert}}]{RCL1994}
{Richichi}, A., {Calamai}, G., \& {Leinert}, C. 1994, \aap, 286, 829

\bibitem[{{R{\"o}ser} {et~al.}(2011){R{\"o}ser}, {Schilbach}, {Piskunov},
  {Kharchenko}, \& {Scholz}}]{SR2011}
{R{\"o}ser}, S., {Schilbach}, E., {Piskunov}, A.~E., {Kharchenko}, N.~V., \&
  {Scholz}, R.-D. 2011, \aap, 531, A92

\bibitem[{{Royer} {et~al.}(2007){Royer}, {Zorec}, \& {G{\'o}mez}}]{RZG2007}
{Royer}, F., {Zorec}, J., \& {G{\'o}mez}, A.~E. 2007, \aap, 463, 671

\bibitem[{{Salaris} \& {Cassisi}(2017)}]{SC2017}
{Salaris}, M., \& {Cassisi}, S. 2017, Royal Society Open Science, 4, 170192

\bibitem[{{Shajn} \& {Struve}(1929)}]{SS1929}
{Shajn}, G., \& {Struve}, O. 1929, \mnras, 89, 222

\bibitem[{{Smart}(1939)}]{WMS1939}
{Smart}, W.~M. 1939, \mnras, 99, 168

\bibitem[{{Soderblom} {et~al.}(2009){Soderblom}, {Laskar}, {Valenti},
  {Stauffer}, \& {Rebull}}]{DS2009}
{Soderblom}, D.~R., {Laskar}, T., {Valenti}, J.~A., {Stauffer}, J.~R., \&
  {Rebull}, L.~M. 2009, \aj, 138, 1292

\bibitem[{{Stauffer} {et~al.}(2007){Stauffer}, {Hartmann}, {Fazio}, {Allen},
  {Patten}, {Lowrance}, {Hurt}, {Rebull}, {Cutri}, {Ramirez}, {Young}, {Rieke},
  {Gorlova}, {Muzerolle}, {Slesnick}, \& {Skrutskie}}]{JRS2007}
{Stauffer}, J.~R., {Hartmann}, L.~W., {Fazio}, G.~G., {et~al.} 2007, \apjs,
  172, 663

\bibitem[{{Tayar} {et~al.}(2017){Tayar}, {Somers}, {Pinsonneault}, {Stello},
  {Mints}, {Johnson}, {Zamora}, {Garc{\'{\i}}a-Hern{\'a}ndez}, {Maraston},
  {Serenelli}, {Allende Prieto}, {Bastien}, {Basu}, {Bird}, {Cohen}, {Cunha},
  {Elsworth}, {Garc{\'{\i}}a}, {Girardi}, {Hekker}, {Holtzman}, {Huber},
  {Mathur}, {M{\'e}sz{\'a}ros}, {Mosser}, {Shetrone}, {Silva Aguirre},
  {Stassun}, {Stringfellow}, {Zasowski}, \& {Roman-Lopes}}]{JT2017}
{Tayar}, J., {Somers}, G., {Pinsonneault}, M.~H., {et~al.} 2017, \apj, 840, 17

\bibitem[{{Taylor}(2006)}]{BT2006}
{Taylor}, B.~J. 2006, \aj, 132, 2453

\bibitem[{{Taylor}(2008)}]{BT2008}
---. 2008, \aj, 136, 1388

\bibitem[{{Taylor} \& {Joner}(2005)}]{TJ2005}
{Taylor}, B.~J., \& {Joner}, M.~D. 2005, \apjs, 159, 100

\bibitem[{{Trampedach} {et~al.}(2015){Trampedach}, {Christensen-Dalsgaard},
  {Asplund}, {Stein}, \& {Nordlund}}]{RT2015}
{Trampedach}, R., {Christensen-Dalsgaard}, J., {Asplund}, M., {Stein}, R.~F.,
  \& {Nordlund}, {\AA}. 2015, in European Physical Journal Web of Conferences,
  Vol. 101, European Physical Journal Web of Conferences, 06064

\bibitem[{{van Altena} {et~al.}(1997){van Altena}, {Lee}, \&
  {Hoffleit}}]{vALH1997}
{van Altena}, W.~F., {Lee}, J.~T., \& {Hoffleit}, E.~D. 1997, Baltic Astronomy,
  6, 27

\bibitem[{{van Belle} {et~al.}(2006){van Belle}, {Ciardi}, {ten Brummelaar},
  {McAlister}, {Ridgway}, {Berger}, {Goldfinger}, {Sturmann}, {Sturmann},
  {Turner}, {Boden}, {Thompson}, \& {Coyne}}]{GvB2006}
{van Belle}, G.~T., {Ciardi}, D.~R., {ten Brummelaar}, T., {et~al.} 2006, \apj,
  637, 494

\bibitem[{{VandenBerg} {et~al.}(2002){VandenBerg}, {Richard}, {Michaud}, \&
  {Richer}}]{VRM2002}
{VandenBerg}, D.~A., {Richard}, O., {Michaud}, G., \& {Richer}, J. 2002, \apj,
  571, 487

\bibitem[{{Viani} {et~al.}(2018){Viani}, {Basu}, {Ong}, {Bonaca}, \&
  {Chaplin}}]{LSV2018}
{Viani}, L.~S., {Basu}, S., {Ong}, J.~J.~M., {Bonaca}, A., \& {Chaplin}, W.~J.
  2018, ArXiv e-prints, arXiv:1803.05924

\bibitem[{{von Zeipel}(1924)}]{HvZ1924}
{von Zeipel}, H. 1924, \mnras, 84, 665

\bibitem[{{Vrancken} {et~al.}(2000){Vrancken}, {Lennon}, {Dufton}, \&
  {Lambert}}]{MV2000}
{Vrancken}, M., {Lennon}, D.~J., {Dufton}, P.~L., \& {Lambert}, D.~L. 2000,
  \aap, 358, 639

\bibitem[{{Wang} {et~al.}(2014){Wang}, {Chen}, {Lin}, {Pandey}, {Huang},
  {Panwar}, {Lee}, {Tsai}, {Tang}, {Goldman}, {Burgett}, {Chambers}, {Draper},
  {Flewelling}, {Grav}, {Heasley}, {Hodapp}, {Huber}, {Jedicke}, {Kaiser},
  {Kudritzki}, {Luppino}, {Lupton}, {Magnier}, {Metcalfe}, {Monet}, {Morgan},
  {Onaka}, {Price}, {Stubbs}, {Sweeney}, {Tonry}, {Wainscoat}, \&
  {Waters}}]{PFW2014}
{Wang}, P.~F., {Chen}, W.~P., {Lin}, C.~C., {et~al.} 2014, \apj, 784, 57

\bibitem[{{Wayman}(1967)}]{PAW1967}
{Wayman}, P.~A. 1967, \pasp, 79, 156

\bibitem[{{Weisz} {et~al.}(2011){Weisz}, {Dalcanton}, {Williams}, {Gilbert},
  {Skillman}, {Seth}, {Dolphin}, {McQuinn}, {Gogarten}, {Holtzman}, {Rosema},
  {Cole}, {Karachentsev}, \& {Zaritsky}}]{DW2011}
{Weisz}, D.~R., {Dalcanton}, J.~J., {Williams}, B.~F., {et~al.} 2011, \apj,
  739, 5

\bibitem[{{Woo} {et~al.}(2003){Woo}, {Gallart}, {Demarque}, {Yi}, \&
  {Zoccali}}]{JHW2003}
{Woo}, J.-H., {Gallart}, C., {Demarque}, P., {Yi}, S., \& {Zoccali}, M. 2003,
  \aj, 125, 754

\bibitem[{{Yen} {et~al.}(2018){Yen}, {Reffert}, {Schilbach}, {R{\"o}ser},
  {Kharchenko}, \& {Piskunov}}]{SXY2018}
{Yen}, S.~X., {Reffert}, S., {Schilbach}, E., {et~al.} 2018, ArXiv e-prints,
  arXiv:1802.04234

\bibitem[{{Zhao} {et~al.}(2009){Zhao}, {Monnier}, {Pedretti}, {Thureau},
  {M{\'e}rand}, {ten Brummelaar}, {McAlister}, {Ridgway}, {Turner}, {Sturmann},
  {Sturmann}, {Goldfinger}, \& {Farrington}}]{MZ2009}
{Zhao}, M., {Monnier}, J.~D., {Pedretti}, E., {et~al.} 2009, \apj, 701, 209

\bibitem[{{Zorec} \& {Royer}(2012)}]{ZR2012}
{Zorec}, J., \& {Royer}, F. 2012, \aap, 537, A120

\bibitem[{{Zwahlen} {et~al.}(2004){Zwahlen}, {North}, {Debernardi}, {Eyer},
  {Galland}, {Groenewegen}, \& {Hummel}}]{NZ2004}
{Zwahlen}, N., {North}, P., {Debernardi}, Y., {et~al.} 2004, \aap, 425, L45

\end{thebibliography}

\newpage

\begin{deluxetable*}{lcccc}[!h]
\centering
\tablewidth{\linewidth}
\tablecaption{Best Fit Ages and [Fe/H]}
\tablehead{ \colhead{Cluster} 
            & \colhead{Filters} 
            & \colhead{$\frac{\Omega}{\Omega_{\rm{c}}}$} 
            & \colhead{Age [Myr]} 
            & \colhead{[Fe/H]}\tablenotemark{1}}
\startdata
The Hyades \\
 & & 0.0 & $776^{+36}_{-15}$ & $0.24^{+0.01}_{-0.03}$ \\
 & $B_T , V_T$ & 0.3 & $676^{+13}_{-30}$ & $0.24^{+0.02}_{-0.01}$ \\
 & & $P(\frac{\Omega}{\Omega_{\rm{c}}} = 0.3)$ & $676^{+67}_{-11}$ & $0.24\pm0.01$ \\ \\
 & & 0.0 & $741^{+17}_{-15}$ & $0.10\pm0.01$ \\
 & $J , K_s$ & 0.6 & $589^{+29}_{-11}$ & $0.12\pm0.01$ \\
 & & $P(\frac{\Omega}{\Omega_{\rm{c}}} = 0.3)$ & $741^{+55}_{-12}$ & $0.10^{+0.01}_{-0.04}$ \\ \hline
The Praesepe \\
 & & 0.0 & $589^{+13}_{-14}$ & $>0.38$ \\
 & $B_T , V_T$ & 0.5 & $589^{+13}_{-26}$ & $0.24^{+0.03}_{-0.02}$ \\
 & & $P(\frac{\Omega}{\Omega_{\rm{c}}} = 0.3)$ & $617^{+40}_{-10}$ & $0.26^{+0.02}_{-0.04}$ \\ \\
 & & 0.0 & $741^{+42}_{-15}$ & $0.08^{+0.01}_{-0.03}$ \\
 & $J , K_s$ & 0.4 & $617^{+14}_{-13}$ & $0.10\pm0.01$ \\
 & & $P(\frac{\Omega}{\Omega_{\rm{c}}} = 0.3)$ & $617^{+17}_{-15}$ & $0.09^{+0.01}_{-0.02}$ \\ \hline
The Pleiades \\
 & & 0.0 & $123^{+3}_{-15}$ & $>0.29$ \\
 & $B_T , V_T$ & 0.6 & $141^{+27}_{-12}$ & $>0.40$ \\
 & & $P(\frac{\Omega}{\Omega_{\rm{c}}} = 0.3)$ & $112^{+2}_{-26}$ & $>0.30$ \\ \\
 & & 0.0 & $162^{+182}_{-65}$ & $>-0.01$ \\
 & $J , K_s$ & 0.6 & $155^{+104}_{-37}$ & $0.18^{+0.29}_{-0.15}$ \\
 & & $P(\frac{\Omega}{\Omega_{\rm{c}}} = 0.3)$ & $155^{+150}_{-46}$ & $>0.03$ \\
\enddata

\tablenotetext{1}{Unbounded values are displayed as lower limits 
                  (e.g., $>0.03$); see the beginning of \S 
                  \ref{s:results}.}

\vspace{0.1cm}
\label{t:results}
\end{deluxetable*}

\end{document}